\documentclass[%
 reprint,
 amsmath,amssymb,
 aps,
nofootinbib,
]{revtex4-2}

\usepackage{graphicx}
\usepackage{dcolumn}
\usepackage{bm}
\usepackage{siunitx}
\usepackage{upgreek}
\usepackage{xfrac}
\usepackage{slashed} 
\usepackage{tikz-feynman}
\usepackage{feynmp-auto}
\usepackage{pgf}
\usepackage{tikz} 
\usetikzlibrary{shapes,arrows,automata}
\tikzfeynmanset{compat=1.1.0}

\def \i {\mathrm{i}}
\def \d {\mathrm{d}}
\def \L {\mathrm{L}}
\def \X {\mathrm{X}}
\def \Y {\mathrm{Y}}
\def \R {\mathrm{R}}
\def \ee {\mathrm{e}}
\def \thetaW {\theta_\mathrm{W}}
\def \cW {\mathrm{c_W}}
\def \sW {\mathrm{s_W}}
\def \s2W {\mathrm{s_W}^2}
\def \tW {\mathrm{t_W}}
\def \Tmax {T_\mathrm{max}}
\def \Trh {T_\mathrm{RH}}
\def \Tds {T^\ast}
\def \rhods {\rho^\ast}
\def \Hrh {H_\mathrm{RH}}
\def \arh {a_\mathrm{RH}}
\def \grh {g_{\rm RH}}
\def \rhorh{\rho_{\rm RH}}
\def \amax {a_\mathrm{max}}
\def \aend {a_\mathrm{end}}
\def \Mp {M_\mathrm{P}}

\def \sigmae {\overline{\sigma}_\mathrm{e}}
\def \Bphi {B_{\phi\chi\chi}}
\def \bea {\begin{eqnarray}}
\def \eea {\end{eqnarray}}
\def \beq {\begin{equation}}
\def \eeq {\end{equation}}

\newcommand{\noi}{\noindent}

\begin{document}

\title{
When direct detection constrains reheating temperature: freeze-in with stronger couplings and inflaton-seeded freeze-in
}

\author{X. Bertou}
\author{O. Deligny}%
\author{M. Gross}%
\author{Y. Mambrini}%
\author{I. Mellouki}%
\affiliation{%
Laboratoire de Physique des 2 Infinis Ir\`ene Joliot-Curie (IJCLab), CNRS/IN2P3, Universit\'{e} Paris-Saclay, Orsay, France}%

\date{\today}

\begin{abstract}
Recent results from the DAMIC-M and PandaX collaborations have excluded the standard freeze-in production of dark matter for masses in the range $3~\mathrm{MeV} \lesssim m_\chi \lesssim 1~\mathrm{GeV}$ in the context of extensions of the Standard Model featuring an additional ultra-light $U(1)_\X$ gauge boson. 
In this work, we analyze the constraints induced by DAMIC-M and PandaX results on the reheating temperature in freeze-in models at stronger coupling, or when a non-thermal source (such as inflaton decay) comes into play.
We identify viable scenarios in which the DM relic abundance is correctly reproduced while evading current experimental bounds on the electron-scattering cross section, $\sigmae$. 
In particular, we show that for reheating temperatures below the electroweak scale, Boltzmann suppressed production can be compensated by stronger couplings, bringing freeze-in scenarios within present experimental reach. 
Finally, we study a hybrid scenario in which a small branching ratio of inflaton decay seeds a nonzero initial dark-matter abundance. 
We show that such contributions can significantly modify freeze-in predictions across broad regions of parameter space, offering an additional pathway for probing extremely feeble interactions. 
\end{abstract}

\maketitle



\section{Introduction}\label{sec:intro}

The nature and origin of dark matter (DM) remain unknown.
As an alternative to the prevailing paradigm of  Weakly Interacting Massive Particle
(WIMP), where DM is in thermal equilibrium in the early Universe and freezes out once the temperature drops below its mass, the freeze-in production 
mechanism~\cite{McDonald:2001vt,Choi:2005vq,Shaposhnikov:2006xi,Kusenko:2006rh,Petraki:2007gq,Hall:2009bx,Bernal:2017kxu} has attracted a lot of 
attention. In this scenario, DM particles never reach thermal equilibrium with the Standard Model (SM) plasma but are instead produced 
non-thermally through feeble interactions with SM particles in the early Universe. This mechanism is particularly well-suited for particle-
physics setups where the interactions of DM are too weak to ensure 
thermalization, typically involving new fields with feeble couplings to 
SM ones, or acting as heavy mediators \cite{Mambrini:2013iaa,Chu:2013jja,Chowdhury:2018tzw,Bernal:2018qlk} like in SO(10) constructions \cite{Mambrini:2016dca,Mambrini:2015vna} or in supergravity framework \cite{Dudas:2017rpa,Benakli:2017whb,Dudas:2017kfz}. 
Emblematic examples of such setups rely on extra-$U(1)$ gauge extensions of the SM, in which DM interacts through a new gauge sector \cite{Lindner:2020kko,Bhattacharyya:2018evo,Dudas:2013sia}.

On the other hand,
light DM candidates can be probed in direct-detection experiments by exploiting inelastic processes involving small internal energy deposits, such as atomic ionization or electronic excitations. These channels are particularly well suited to sub-GeV DM, for which nuclear recoils fall below threshold but electron recoils remain observable. The sensitivity is further enhanced in scenarios with ultra-light mediators, where the scattering cross section exhibits a pronounced enhancement at low momentum transfer. As a result, the recoil spectrum is sharply peaked toward small recoil energies, leading to a substantial increase in the total event rate. This feature allows direct-detection experiments to probe extremely small interaction strengths, including the feeble couplings characteristic of freeze-in scenarios. 

Recent results from the DAMIC-M and PandaX collaborations have demonstrated the growing impact of low-threshold electron-recoil searches on this class of models. In particular, DAMIC-M has set stringent limits on ultra-weak DM–electron interactions for DM masses in the MeV to sub-GeV range~\cite{DAMIC-M:2025luv}, while PandaX has extended sensitivity to masses in the multi-GeV range~\cite{PandaX:2025rrz}. These results significantly constrain the parameter space of freeze-in DM in the benchmark scenario featuring an additional $U(1)$ gauge boson kinetically mixed with the SM hypercharge~\cite{Chu:2011be,Essig:2011nj,Bhattiprolu:2023akk}. In this framework, the viable region in the plane of the DM–electron scattering cross section $\sigmae$ and DM mass $m_\chi$ has been substantially reduced. Given the projected improvements in detector exposure and threshold sensitivity in upcoming experimental runs, it is timely to further develop the phenomenology of freeze-in DM and to delineate the range of theoretical possibilities that may be explored in the near future for experiments sensitive to ultra-light mediators.

However, in the case of FIMP-type models, 
it is often necessary 
to justify the 
absence of initial conditions—since 
thermal equilibrium 
is never reached—as 
well as the feebly coupled nature of 
dark matter with the thermal bath. A 
compelling proposal that has recently 
emerged is called Freeze-In at Stronger Coupling (FISC) \cite{Cosme:2023xpa}, 
where the authors assume a very low 
reheating temperature $\Trh$. Indeed, apart 
from the constraint from Big Bang 
Nucleosynthesis (BBN), which imposes 
a value $\Trh \gtrsim 4 \text{ MeV}$, there is no experimental 
data or theoretical argument that forbids 
having reheating temperatures well 
below the electroweak scale, or even down 
to a few tens of MeV.

The advantage of such a framework is that 
it is possible to ``transfer'' the burden 
of the coupling's weakness onto the 
Boltzmann suppression factor for the 
production of dark matter with a mass 
$m_\chi \gg \Trh$. Dark matter can then 
possess couplings of the order of the 
electroweak scale (thereby recovering 
the motivations of unified models), 
making them ``effectively'' FIMPs while their coupling 
remains standard, and thus observable\footnote{Notice also that a very low reheating 
temperature considerably reduces 
one of the most dangerous (because 
unavoidable) sources of initial 
conditions, namely the one arising from 
graviton exchange in the thermal bath \cite{Clery:2021bwz}.
}. Under these 
conditions, direct detection experiments are no longer probes 
of couplings, but rather of the reheating 
temperature, transforming them 
into ``cosmic explorers.''

In this work, we investigate in detail 
the constraints imposed by direct detection experiments 
within a freeze-in scenario at a low reheating 
temperature. In particular, we show how the DAMIC-M 
experiment is already capable of considerably reducing 
the viable cosmological parameter space, 
notably establishing a lower bound on the reheating temperature 
$\Trh \gtrsim 1 \text{GeV}$, which is stronger than the BBN bound. Furthermore, we aim to relax the 
constraint on initial conditions by accounting for a non-
thermal production source via the decay of the inflaton field 
$\phi$ into the dark sector, represented by the dark matter candidate $\chi$. 
Indeed, unless one considers the inflaton responsible 
for reheating to be ``dark-matter-phobic''
(thereby requiring a charged inflationary sector), it is 
impossible to neglect this initial production channel 
even before the end of reheating. Parameterizing our 
ignorance of the inflaton-dark matter coupling through the 
branching ratio $\text{BR}_{\phi\rightarrow \chi \chi}$, we 
analyze the constraints obtained on this branching 
ratio using DAMIC-M data. Finally, we propose to extend our 
study by projecting the future sensitivities of 
DAMIC-M.

The structure of this paper is as follows. 
In Section~\ref{sec:DS}, we review briefly the possibilities to extend the electroweak theory with an extra-$U(1)_\X$ gauge symmetry. 
We focus on representative scenarios that either involve kinetic mixing with the SM hypercharge--allowing communication with an otherwise secluded dark sector--or make use of a global symmetry above the electroweak scale that is gauged and embedded within the dark sector. 
In Section~\ref{sec:freeze-in}, we review the freeze-in production mechanism of DM in the case of low-reheating temperatures and discuss the current and projected experimental probes of $\sigmae$ as a function of $m_\chi$.  
It is shown in particular that, for low-enough reheating temperatures, the vanilla freeze-in scenario can enter into a regime of freeze-in at stronger coupling that improves the detection prospects. 
In Section~\ref{sec:infdecay}, we consider the contribution from inflaton (or reheaton) decays as a concrete example of DM production {\it before} the end of reheating, 
addressing the initial condition issue. 
We demonstrate that, across broad ranges of reheating temperatures and inflaton branching ratios into DM, the resulting nonzero initial abundance can significantly impact the freeze-in yield. 
This scenario offers a promising pathway for probing extremely feeble DM interactions with both current and upcoming direct-detection experiments. 
Finally, our main results and conclusions are summarized in Section~\ref{sec:conclusion}.

\section{Dark-sector model}\label{sec:DS}

\subsection{$U(1)$ extensions of electroweak interactions}\label{sec:U1}

One of the most conservative approaches to extend the SM consists in introducing a new spin-1 mediator coupled to a gauge-symmetry current associated with conserved quantum numbers. Interestingly, above the electroweak scale, there are several unbroken colorless $U(1)$ symmetry generators that can be combined into an extra-hypercharge linearly independent of the SM one. The extra-$U(1)_\X$ hypercharge, denoted as $X$, is known to take the general form~\cite{Fayet:1989mq}
\begin{equation} \label{Eq:X}
    X=bB+\ell_i L_i+yY+Q_X,
\end{equation}
where $B$ stands for the baryonic number, $L_i$ the leptonic ones for each flavor, $Y$ the SM hypercharge, and $Q_X$ an additional quantum number related to a dark sector. In this context, the extended electroweak covariant derivative takes the form
\begin{equation} \label{Eq:Dmu}
    D_\mu =\partial_\mu-\i g\mathbf{T}\cdot\mathbf{W}_\mu-\i\frac{g'}{2}YB_\mu-\i\frac{g''}{2}XC_\mu,
\end{equation}
with $(g,g',g'')$ the coupling constants of the gauge group $SU(2)_\L\times U(1)_\Y\times U(1)_\X$, $(\mathbf{T},Y,X)$ the corresponding generators, and $(\mathbf{W}_\mu,B_\mu,C_\mu)$ the vector-boson mediators. 

In addition to nonzero $b$, $\ell_i$ or $y$ values, feeble interactions between the SM and dark sectors can also occur through kinetic mixing $\epsilon$ between the field-strength tensors $B_{\mu\nu}$ and $C_{\mu\nu}$ associated with the gauge fields $B_\mu$ and $C_\mu$, respectively~\cite{Holdom:1985ag},
\begin{equation}
\label{Eq:L_gauge}
\mathcal{L}_\mathrm{gauge}
=-\frac{1}{4}B_{\mu\nu}B^{\mu\nu}
-\frac{1}{4}C_{\mu\nu}C^{\mu\nu}
-\frac{\epsilon}{2}B_{\mu\nu}C^{\mu\nu}.
\end{equation}
Such kinetic mixing can naturally arise from a sector of heavy fermionic messenger fields charged under both the SM hypercharge $Y$ and the dark charge $Q_X$. After integrating out these states, the mixing term is generated in the low-energy effective theory.

The dark sector is assumed to be composed of Dirac fermions denoted as $\chi$ that are charged under $U(1)_\X$ and couple to the $C_\mu$ boson in the interaction basis,
\begin{equation}
\label{Eq:L_DS}
\mathcal{L}_\mathrm{DS}=\overline{\chi}(\i\slashed \partial-m_\chi)\chi-\frac{m^2_C}{2}C_\mu C^\mu+\frac{g''}{2}Q_X\overline{\chi}\gamma^\mu\chi C_\mu,
\end{equation}

\noi
where the mass term $m_C$ is assumed to range below the keV scale so that the regime of ``ultra-light mediator'' is fulfilled. Note that $C_\mu$ may even remain massless for most of the studies presented below. 

Overall, depending on the choices of the parameters $b$, $\ell_i$, $y$, and $\epsilon$, feeble interactions between the SM and dark sectors can be realized in several distinct ways. 
Interestingly, models with $X=Q_X$ and a nonzero kinetic mixing are poorly constrained in terms of $g''$ and $\epsilon$ for a wide range of $m_{A'}$ below the keV scale.\footnote{On the other hand, models in which SM particles carry a nonzero charge $X$ are subject to the stringent constraint ($g'' \lesssim 10^{-24}$), arising from equivalence-principle tests that probe the long-range force induced between ordinary bodies~\cite{Fayet:2017pdp,Fayet:2018cjy}. Such a bound is prohibitively strong for most dark-sector realizations relevant to DM phenomenology. 
In addition, for models with $X = L_\mu - L_\tau + Q_X$, constraints from white-dwarf luminosity functions imply $g'' \lesssim 10^{-6}$~\cite{Alonso-Gonzalez:2025xqg}. This bound is borderline for viable dark-sector realizations of DM phenomenology.}
We thus restrict ourselves to such a realization below, outlining the main features of the model, whereas technical details are deferred to Appendix~\ref{app:models_withepsilon}.

\subsection{Benchmark model}

As a benchmark, we adopt a choice widely used in the 
literature, namely $b=\ell_i=y=0$ with a nonvanishing kinetic 
mixing $\epsilon$ between 
the field-strength tensors $B_{\mu\nu}$ and $C_{\mu\nu}$. In this case, the gauge fields 
must be redefined to 
eliminate the mixed term $-\epsilon B_{\mu\nu}C^{\mu\nu}/2$ from the kinetic 
Lagrangian before expressing the covariant 
derivative in terms of canonically normalized fields 
$(\hat{W}_{3\mu},\hat{B}_\mu,\hat{C}_\mu)$. This field redefinition leaves 
one free mixing angle. In the present benchmark, we 
choose it such that the field $\hat{C}_\mu$ alone 
couples to both the SM and the dark sectors. Other setups with different 
choices of mixing angle and extra-hypercharge are 
presented in Appendix~\ref{app:models_withepsilon}.

To first order in $\epsilon$, the field  transformation in this setup reads as
\begin{equation}
\label{Eq:nonhat2hat}
    \begin{pmatrix}
        W_{3\mu}\\
        B_\mu\\
        C_\mu
    \end{pmatrix}=\begin{pmatrix}
        \hat{W}_{3\mu}\\
        \hat{B}_\mu-\epsilon\hat{C}_\mu\\
        \hat{C}_\mu
    \end{pmatrix},
\end{equation}
which leads to the covariant derivative 
\begin{equation}\label{Eq:Dmu}
    D_\mu=\partial_\mu-\mathrm{i}gT_3\hat{W}_{3\mu}-\frac{\mathrm{i}}{2}g'Y\hat{B}_\mu
    -\frac{\mathrm{i}}{2}\left(g''Q_X-\epsilon g'Y\right)\hat{C}_\mu.
\end{equation}
At temperatures sufficiently high for electroweak interactions to remain in the symmetric phase, DM production from the thermal bath is then mediated by the diagram
\begin{equation}\label{Eq:diagram-1}
\centering
\begin{tikzpicture}
\node at (2.5,-0.6) {\( \hat{C}_\mu \)};
\node at (0.3,-1) {\( \i\frac{\epsilon g'}{2}Y_{f_{\L/\R}}\gamma^\mu \)};
\node at (4.6,-1) {\( -\i \frac{g''}{2}\gamma^\mu, \)};
\begin{feynman}
    \vertex (a1){\(f_{\L/\R}\)};
    \vertex [right=1.5cm of a1] (a2);
    \vertex [below=1cm of a2] (a3);
    \vertex [below=2cm of a1] (a4){\(\overline{f}_{\L/\R}\)};
    \vertex [right=1.5cm of a4] (a5); 
    \vertex [right=2cm of a3] (a6); 
    \vertex [right=1.5cm of a6] (a7); 
    \vertex [above=0.8cm of a7] (a8){\(\chi\)};
    \vertex [below=2cm of a8] (a9){\(\overline{\chi}\)};
    \diagram*{
      {[edges=fermion] (a1) --[plain] (a3)},  
      {[edges=anti fermion] (a4) --[plain] (a3)},
      {(a3) --[photon, thick] (a6)},
      {[edges=fermion] (a6) --[plain] (a8)},  
      {[edges=anti fermion] (a6) --[plain] (a9)},  
    };
    \draw [fill=black] (a3) circle (2pt);
    \draw [fill=black] (a6) circle (2pt);
  \end{feynman}
\end{tikzpicture}
\nonumber
\end{equation}
with $Y_{f_{\L/\R}}$ the hypercharge of the SM fermion $f_{\L/\R}$ with left or right chirality 
and $Q_X^\chi=1$. On the 
other hand, for temperatures below the 
electroweak transition, the  three physical gauge 
bosons are the eigenstates of the mass-squared matrix. We show in Appendix \ref{app:xs} that they are obtained from the hatted fields by $3\times 3$ rotations around $\hat{B}_\mu$ and $\hat{C}_\mu$:
\begin{equation}\label{Eq:AZA'}
    \begin{pmatrix}
        \tilde{Z}_\mu\\A_\mu\\A'_\mu
    \end{pmatrix}=\mathcal{R}_{\hat{B}_\mu}(+\xi)\cdot \mathcal{R}_{\hat{C}_\mu}(+\thetaW)\begin{pmatrix}
        \hat{W}_{3\mu}\\\hat{B}_\mu\\\hat{C}_\mu
    \end{pmatrix},
\end{equation}
with $\thetaW$ the electroweak mixing angle and $\xi\simeq\epsilon\sin\thetaW$. In this particular electroweak extension, the $A_\mu$ boson is the standard photon and the $A_\mu'$ boson is the ``dark photon''. The third physical gauge bosons, $\tilde{Z}_\mu$, is a modified version of the SM $Z_\mu$ boson, with a mass receiving corrections to order $\epsilon^2$. The DM production is then governed by two interfering diagrams, one with a dark photon mediator,
\begin{equation}\label{Eq:diagram-2}
\centering
\begin{tikzpicture}
\node at (2.5,-0.6) {\( A'_\mu \)};
\node at (0,-1) {\( \i\epsilon e Q_f\cW\gamma^\mu \)};
\node at (4.6,-1) {\( -\i \frac{g''}{2}\gamma^\mu, \)};
\begin{feynman}
    \vertex (a1){\(f\)};
    \vertex [right=1.5cm of a1] (a2);
    \vertex [below=1cm of a2] (a3);
    \vertex [below=2cm of a1] (a4){\(\overline{f}\)};
    \vertex [right=1.5cm of a4] (a5); 
    \vertex [right=2cm of a3] (a6); 
    \vertex [right=1.5cm of a6] (a7); 
    \vertex [above=0.8cm of a7] (a8){\(\chi\)};
    \vertex [below=2cm of a8] (a9){\(\overline{\chi}\)};
    \diagram*{
      {[edges=fermion] (a1) --[plain] (a3)},  
      {[edges=anti fermion] (a4) --[plain] (a3)},
      {(a3) --[photon, thick] (a6)},
      {[edges=fermion] (a6) --[plain] (a8)},  
      {[edges=anti fermion] (a6) --[plain] (a9)},  
    };
    \draw [fill=black] (a3) circle (2pt);
    \draw [fill=black] (a6) circle (2pt);
  \end{feynman}
\end{tikzpicture}
\nonumber
\end{equation}
and another one with the modified $Z$ boson mediator,
\begin{equation}\label{Eq:diagram-3}
\centering
\begin{tikzpicture}
\node at (2.5,-0.6) {\( \tilde{Z}_\mu \)};
\node at (-1,-1) {\( -\i\frac{e}{2\sW\cW} (v_{f}-a_{f}\gamma^5)\gamma^\mu\)};
\node at (4.6,-1) {\( -\i\frac{g''\epsilon \sW}{2}\gamma^\mu, \)};
\begin{feynman}
    \vertex (a1){\(f\)};
    \vertex [right=1.5cm of a1] (a2);
    \vertex [below=1cm of a2, blob, minimum size=8pt] (a3);
    \vertex [below=2cm of a1] (a4){\(\overline{f}\)};
    \vertex [right=1.5cm of a4] (a5); 
    \vertex [right=2cm of a3] (a6); 
    \vertex [right=1.5cm of a6] (a7); 
    \vertex [above=0.8cm of a7] (a8){\(\chi\)};
    \vertex [below=2cm of a8] (a9){\(\overline{\chi}\)};
    \diagram*{
      {[edges=fermion] (a1) --[plain] (a3)},  
      {[edges=anti fermion] (a4) --[plain] (a3)},
      {(a3) --[photon, thick] (a6)},
      {[edges=fermion] (a6) --[plain] (a8)},  
      {[edges=anti fermion] (a6) --[plain] (a9)},  
    };
    \draw [fill=black] (a3) circle (2pt);
    \draw [fill=black] (a6) circle (2pt);
  \end{feynman}
\end{tikzpicture}
\nonumber
\end{equation}
with $v_f$ and $a_f$ the vector and axial couplings associated between the SM $Z_\mu$ boson each fermion. 

The cross section for direct-detection, $\sigmae$, follows from the DM-electron scattering in the $t$-channel. In the kinematic range of interest, and considering the leading order contribution from the $A'_\mu$ only, $\sigmae$ reads as \cite{DAMIC-M:2025luv,Mambrini:2021cwd}

\begin{equation}\label{Eq:sigma_e}
    \sigmae\simeq \frac{(\epsilon eg''\cW)^2}{4\pi}\frac{\mu^2_{\mathrm{e}\chi}}{(\alpha^2 m_\mathrm{e}^2+m_{A'}^2)^2},
\end{equation}

\noi
where

\beq
\mu_{\mathrm{e}\chi}=\frac{m_\chi m_\mathrm{e}}{(m_\chi+m_\mathrm{e})}
\eeq

\noi
is the reduced mass and $\alpha=1/137$ is the fine structure constant. We have made use of the fact that the typical momentum transfer in the reaction is $q\simeq\alpha m_\mathrm{e}$.

\section{Dark matter relic abundance }
\label{sec:freeze-in}

\subsection{The freeze-in at stronger coupling mechanism}

The current tension in the search for WIMP candidates arises from the requirements of the standard thermal freeze-out mechanism during the radiation domination era. 
Achieving the observed relic abundance calls for a sufficiently large coupling between the DM particle $\chi$ and the thermal bath to deplete the relic yield down to the required level, namely $n_\chi^0 / n_\gamma^0 \lesssim 10^{-9}$ for $m_\chi \sim \mathcal{O}(1\,\mathrm{GeV})$. 
In contrast, the FIMP paradigm relies on extremely small couplings to avoid thermal equilibrium and overproduction of DM. 
This makes the detection even more challenging. 

The scenario proposed in Ref.~\cite{Cosme:2023xpa} considers instead a cosmological history in which the thermal bath never reaches high temperatures, remaining below the electroweak scale, or even at the GeV scale. 
Importantly, the only robust lower bound on the reheating temperature is set by primordial nucleosynthesis (BBN), namely $T_{\mathrm{RH}} \gtrsim 4\,\mathrm{MeV}$~\cite{Sarkar:1995dd,Kawasaki:1999na,Hannestad:2004px,deSalas:2015glj}. 
In such a low-reheating-temperature framework, the suppression typically attributed to feeble couplings can instead be transferred to the Boltzmann factor, $\exp(-m_\chi / T_{\mathrm{RH}})$, provided $m_\chi \lesssim \mathcal{O}(10)\,T_{\mathrm{RH}}$.\footnote{A more precise numerical factor depends on the production channel and cosmological history.} 
This mechanism, referred to as ``freeze-in at stronger coupling'' (FISC), retains the advantages of the FIMP framework while enhancing the detection prospects due to interaction strengths large enough to yield observable signals in both collider and direct detection experiments. 
Interestingly, since the low reheating temperature truncates the thermal history and pushes production toward the highest available temperature, DM production is UV sensitive in contrast to the ``vanilla'' freeze-in framework built upon the $U(1)$ extension considered in this study.

DM production is then governed by the Boltzmann equation
\begin{equation}
\frac{dn_\chi}{dt} + 3 H n_\chi = R(T),
\label{Eq:boltzmannt}
\end{equation}
where $T$ denotes the temperature of the thermal bath and $H$ the Hubble expansion rate. 
The latter is set by the dominant energy component during the freeze-in epoch, which may be either the inflaton field or radiation. 
In the following, we first consider the limit of instantaneous reheating, such that the subsequent evolution corresponds to a purely thermal, radiation-dominated Universe, from the reheating temperature $T_{\rm RH}$ down to temperatures relevant for DM production.

It is convenient to recast Eq.~(\ref{Eq:boltzmannt}) in terms of the scale factor $a$ by introducing the comoving number density $X_\chi \equiv n_\chi a^3$. 
This yields
\begin{equation}
\frac{dX_\chi}{da} = \frac{a^2 R(T)}{H}.
\label{Eq:boltzmanna1}
\end{equation}
Using the relation between temperature and scale factor during radiation domination,
\begin{equation}
T = T_{\rm RH} \left(\frac{g_{\rm RH}}{g_\star}\right)^{1/3} \frac{a_{\rm RH}}{a},
\end{equation}
together with $H = \sqrt{\rho/(3 \Mp^2)}$ and $\rho = (\pi^2/30) g_\star T^4 \equiv \tilde{\alpha} T^4$ with $g_\star$ the effective number of relativistic degrees of freedom at temperature $T$, one obtains
\begin{equation}
\frac{dX_\chi}{da} = \frac{a^4}{a_{\rm RH}^2}
\sqrt{\frac{3}{\tilde{\alpha}}}
\frac{\Mp}{T_{\rm RH}^2}
\left(\frac{g_\star}{g_{\rm RH}}\right)^{2/3}
R(T),
\label{Eq:boltzmanna}
\end{equation}
with $\Mp\simeq 2.4\times10^{18}~$GeV the reduced Planck mass.
The remaining task is therefore to compute the production rate $R(T)$ and express it as a function of the scale factor $a$.

\subsection{The production rate}

The production rate for a $1+2 \to 3+4$ process, describing annihilation into DM via $i i \to \chi \chi$, can be written in the general form
\begin{multline}
\label{eqn:rate}
R(T) = \frac{1}{2048 \pi^6} \int f_1 f_2 \, E_1 \, dE_1 \, E_2 \, dE_2 \, d\cos\theta_{12} \\
\times \int |{\cal M}|^2  \beta_i(s)\beta_\chi(s)\, d\Omega_{13},
\end{multline}
where $\beta_i(x) \equiv \sqrt{1 - 4 m_i^2/x}$, $\Omega_{13}$ denotes the solid angle between particles 1 and 3, and ${\cal M}$ is the invariant amplitude for the process $1+2 \to 3+4$ with center-of-mass energy $s$. 
The distribution functions $f_i$ correspond to Bose–Einstein or Fermi–Dirac statistics, 

\beq
f_i=\frac{g_i}{e^{\frac{E_i}{T}}\pm 1}\,,
\eeq

\noi
which include here the internal degrees of freedom $g_i$ of the particles $i$. 
All species in the thermal bath contribute to DM production. 
While the results we will display in our figures are obtained from a complete numerical evaluation of Eq.~\eqref{eqn:rate} 
(see Appendix~\ref{app:numerical}), 
we also derive approximate analytical expressions below to highlight the main dependencies of the production rate and the resulting relic abundance.

As we have seen, in the case of the FISC-type model, it is the Boltzmann suppression due to a mass $m_\chi > T$ that ensures the weakness of the coupling, thereby avoiding an overabundance of DM. Neglecting the initial particles masses (in the thermal bath), {\it but} considering the final particle is heavier than the bath temperature, the rate is then given by
\begin{multline}
R(T) = \frac{1}{2048 \pi^6} \int f_1 f_2 \, E_1 \, dE_1 \, E_2 \, dE_2 \, d\cos\theta_{12}
\\
\times
\int |{\cal M}|^2 \beta_\chi(s)\, d\Omega_{13} .
\end{multline}

\noi
After the change of variables $(E_1,E_2,\cos\theta_{12}) \to (E_+=E_1+E_2, E_-=E_1-E_2, s)$, the rate becomes

\begin{multline}
R(T) = \frac{g_1g_2T}{1024 \pi^5} \int_{4m_\chi^2}^\infty ds \, K_1(\sqrt{s}/T) \sqrt{s}\beta_\chi(s) \\
\times \int |{\cal M}|^2 \, d\cos\theta_{13},
\end{multline}

\noi
where we have used the approximation $f_1 f_2 \simeq  g_1g_2e^{-E_+/T}$, with $g_1g_2=4$ in the following as the dominant production is from SM fermions. This leads, under the assumption $s > 4m_\chi^2 \gg T$ and after integration over $\Omega_{13}$, to

\begin{equation}
R(T) \simeq \sqrt{\frac{\pi}{2}} \frac{T^{3/2}}{128 \pi^5} \int_{4m_\chi^2}^\infty ds\,s^{1/4} \ee^{-\sqrt{s}/T} \beta_\chi(s)\, |{\cal M}|^2\,.
\end{equation}
Expanding around $s = 4m^2_\chi + \epsilon$ and retaining only the dominant terms in the polynomial expansions under the condition $\epsilon/(4m_\chi^2)\ll 1$, we find
\begin{equation}
R(T) \simeq \frac{ \sqrt{\pi}\, m_\chi T^3}{32 \pi^5} \, \ee^{-\frac{2m_\chi}{T}} \int_0^\infty \sqrt{x} \, \ee^{-x} \, dx \, |{\cal M}|^2,
\end{equation}

\noi
with $x = \epsilon/(4 m_\chi T)$. Assuming a constant squared matrix element $|{\cal M}|^2 \sim g_\chi^2 g_{\rm SM}^2$, this simplifies to

\bea
R(T) 
\simeq  g_\chi^2 g_{\rm SM}^2 \frac{ m_\chi T^3}{64 \pi^4} \,  \ee^{-\frac{2m_\chi}{T}},
\label{Eq:r(t)}
\eea

\noi
where we have used $\Gamma(3/2) = \sqrt{\pi}/2$. Substituting $T = T_{\rm RH} \left(\frac{\grh}{g_\star}\right)^\frac13\frac{ a_{\rm RH}}{a}$ into Eq.~\eqref{Eq:r(t)}, we can now solve Eq.~\eqref{Eq:boltzmanna}.

\subsection{The relic abundance}

We now have all the ingredients required to compute the relic abundance of $\chi$. 
Indeed, Eq.~\eqref{Eq:boltzmanna} can now be written as

\begin{equation}
\frac{dX_\chi}{da} = a^2 \left(\frac{a_{\rm RH}}{a}\right) 
\left(\frac{g_{\rm RH}}{g_\star}\right)^{\frac13} 
\sqrt{\frac{3}{\tilde \alpha}}
\frac{\Mp m_\chi T_{\rm RH}}{64 \pi^4} \, 
\ee^{-\frac{2 m_\chi}{T}} \, |{\cal M}|^2 \,,
\end{equation}

\noi
which, after integration, gives
\begin{multline}
X_\chi(a)  = 
a_{\rm RH} 
\left(\frac{g_{\rm RH}}{g_\star}\right)^{\frac13} 
\sqrt{\frac{3}{\tilde \alpha}} \frac{\Mp m_\chi T_{\rm RH}}{64 \pi^4} \, |{\cal M}|^2 \\
\times 
\int_{\arh}^\infty \ee^{-\frac{2 m_\chi}{\Trh}\frac{a'}{\arh}
\left (\frac{g_\star}{g_{\rm RH}}\right)^\frac13} a' \, da'
\,,
\label{Eq:Xlowrh}
\end{multline}

\noi
where we assumed negligible production during reheating, i.e.\ $X_\chi(a_{\rm RH})=0$ and $\cal M$ independent of $T$ (and thus of $a$), which is our case of study. We deduce
\begin{multline}
X_\chi(a) = 
a_{\rm RH}^3 \frac{g_{\rm RH}}{g_\star}
\sqrt{\frac{3}{\tilde \alpha}} \frac{\Mp T_{\rm RH}^3}{m_\chi} 
\frac{|{\cal M}|^2}{256 \pi^4} 
\\
\times
\int_{\frac{2m_\chi}{T_{\rm RH}} \left(\frac{g_\star}{g_{\rm RH}}\right)^{\frac13}}^\infty 
\ee^{-x} \, x \, dx \,.
\end{multline}
\noi
Approximating the integral 
$\int_a^\infty \ee^{-x} x \, dx \simeq (1+a) \ee^{-a}$, we then obtain

\begin{multline}
n^{\rm FI}(a) \simeq 
\frac{g_{\rm RH}}{g_\star} \left(\frac{a_{\rm RH}}{a}\right)^3 
\sqrt{\frac{3}{\tilde \alpha}} \frac{\Mp}{m_\chi} 
\frac{|{\cal M}|^2}{256 \pi^4} T_{\rm RH}^3 
\\
\times
\left(1 + 2\frac{m_\chi}{T_{\rm RH}} \left(\frac{g_\star}{g_{\rm RH}}\right)^{\frac13}\right) 
\,\ee^{-\frac{2m_\chi}{T_{\rm RH}}}\,,
\end{multline}

\noindent
or equivalently

\begin{multline}
n^{\rm FI}(a) \simeq \sqrt{\frac{3}{\tilde \alpha}} 
\frac{\Mp}{m_\chi} 
\frac{|{\cal M}|^2}{256 \pi^4} T^3 \\
\times
\left(1 + 2\frac{m_\chi}{T_{\rm RH}} 
\left(\frac{g_\star}{g_{\rm RH}}\right)^{\frac13}\right) 
\,\ee^{-\frac{2m_\chi}{T_{\rm RH}}}\,.
\end{multline}

\noi
This leads to the relic density \cite{Mambrini:2021cwd}

\begin{equation}
\Omega_\chi h^2 = 1.6 \times 10^8 \left(\frac{g_0}{g_{\rm RH}}\right) 
\left(\frac{n(\Trh)}{T_{\rm RH}^3}\right)  \left(\frac{m_\chi}{1\,\text{GeV}}\right) \,,
\label{Eq:omegageneric}
\end{equation}

\noi
or

\begin{equation}
    \Omega_\chi^{\rm FI} h^2 \simeq 2.25\times 10^{21}\left(\frac{18}{g_{\rm RH}}\right)
      \left|{\cal M}\right|^2
    \left(1 + \frac{2m_\chi}{T_{\rm RH}} 
    \right) 
    \ee^{-\frac{2m_\chi}{T_{\rm RH}}}\,,
\label{Eq:omegatotal}
\end{equation}

\noi
where we took $g_0=3.91$, $\grh=g_\star$, and the reference value $\grh=18$ corresponding to a reference $\Trh=100~$MeV. 
In the case where $m_\chi \gg T_{\rm RH}$, we obtain

\begin{equation}
\frac{\Omega_\chi^{\rm FI} h^2}{0.12} \simeq 
3.75\times 10^{22} \left(\frac{18}{g_{\rm RH}}\right) \frac{m_\chi}{\Trh}
~|{\cal M}|^2 \ee^{-\frac{2m_\chi}{T_{\rm RH}}}\,.
\label{Eq:omega2}
\end{equation}

Before going further, it is interesting to examine Eq.~\eqref{Eq:omega2} in more detail. 
The numerical prefactor, of order $10^{22}$, may seem enormous at first sight. 
However, the Boltzmann suppression guarantees an acceptable relic density provided that 
$m_\chi \simeq 24 \, T_{\rm RH}$ 
and $|{\cal M}| \simeq 1$. 
This is the key advantage of FISC models, in which the correct relic density can be achieved 
thanks to a difference of only about one order of magnitude between the reheating temperature 
and the DM mass. 
Another interesting point is the UV nature of FISC-type FIMP models. Indeed, as Boltzmann suppression is weakest at high temperatures, all of the DM is produced at the very start of the production process, at $T =\Trh$. 
This also justifies setting $g_\star=\grh$.
It is now time to relate the interaction amplitude ${\cal M}$ to the direct detection rate.

\subsection{The direct detection constraints}

Starting from Eq.~\eqref{Eq:sigma_e} and applying ${\cal M} \sim g_\chi g_{\rm SM} \simeq \epsilon eg''\cW$, 
we can express the relic density, Eq.~\eqref{Eq:omega2} in terms of the cross section $\bar{\sigma}_e$. 
For $m_{A'} \ll m_e$, we obtain
\begin{equation}
|{\cal M}|^2 \simeq 4 \pi \alpha^4 \left(\frac{m_e}{m_\chi}\right)^2 (m_e + m_\chi)^2 ~\bar \sigma_e\,,
\end{equation}
\noindent
which, upon substitution into Eq.~(\ref{Eq:omegatotal}), gives

\begin{multline}
\frac{\bar{\sigma}_e}{3 \times 10^{-37}~\rm cm^2} \simeq \left[1+2\left(\frac{100~\rm MeV}{\Trh} \right)\left(\frac{m_\chi}{1~\rm GeV}\right)\right]^{-1}\\
\times \left(\frac{\Omega_\chi^{\rm FI} h^2}{0.12}\right)
\ee^{\frac{2m_\chi}{T_{\rm RH}}} \,,
\label{Eq:omega3}
\end{multline}

\noindent
where we have assumed $m_\chi \gg m_e$.

\begin{figure}[t]
\centering
\includegraphics[width=0.5\textwidth]{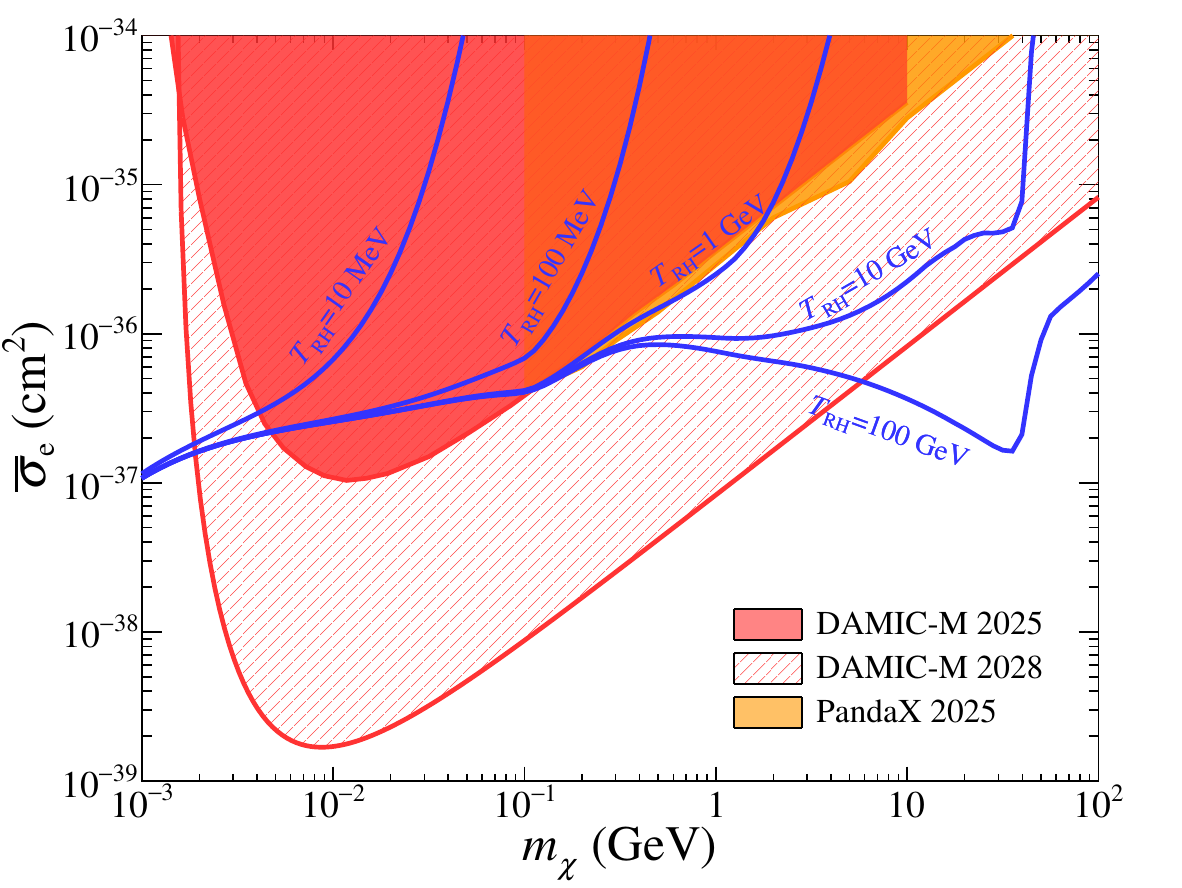} 
\caption{DM-electron cross section as a function of $m_\chi$ for $\Trh$ ranging from 10~MeV to 100~GeV.  The blue curves correspond to the iso-relic density for the specified reheating temperature.}
\label{fig:xs_lowTrh}
\end{figure}

We present in Fig.~\ref{fig:xs_lowTrh} the regions of parameter space satisfying $\Omega_\chi^{\rm FI} h^2 = 0.12$ in the $(m_\chi, \bar{\sigma}_e)$ plane. 
These results are obtained from a numerical solution of the Boltzmann equation, for reheating temperatures in the range $10~\mathrm{MeV} \le \Trh \le 100~\mathrm{GeV}$. 
Iso-relic-density contours corresponding to different values of $T_{\rm RH}$ are shown, together with the current limits from the 2025 DAMIC-M and PandaX experiments~\cite{DAMIC-M:2025luv,PandaX:2025rrz}.
The analytical expression derived previously turns out  to reproduce the results quite accurately, except when the resonance effects of the $Z$ boson enter into play and change substantially $|\mathcal{M}|^2$, at $m_\chi \simeq \frac12 M_Z$. 
We find that, independently of the reheating temperature, the freeze-in realization of the benchmark model is excluded in the mass range $5 \lesssim (m_\chi/\mathrm{MeV}) \lesssim 300$. 
The projected sensitivity for DAMIC-M is also displayed as the shaded region, showing that the parameter space corresponding to higher reheating temperatures, up to $T_{\rm RH} \sim 50~\mathrm{GeV}$, may be probed in the next two years. 
The projected sensitivity is derived using the same profile-likelihood framework as that reported in Ref.~\cite{DAMIC-M:2025luv}, where the expected event rate for each candidate pattern is modeled as the sum of (i) single-electron background events, estimated from the measured rate of 1-electron pixels, and (ii) radiogenic backgrounds, the contribution of which is obtained from \textsc{Geant4}-based simulations and normalized to the observed number of higher-energy events in a control region. 
For the projection, the exposure is scaled by two orders of magnitude, assuming unchanged detector performance and ten-times reduced background rates. 
The resulting sensitivity improvement lies between the linear and square-root scaling with exposure expected in the background-free and background-limited regimes, respectively. 
Over the same timescale, improvement in sensitivity is also expected for PandaX scaled with 20T, but, as the current search is background limited, the exact reach will depend on the level of background reduction ultimately obtained. 
With optimistic assumptions, a sensitivity comparable to that projected for DAMIC-M appears achievable.

It is also interesting to note the ``Boltzmann factor'' effect, which tends to strongly increase the direct detection cross section $\bar{\sigma}_e$ for a fixed relic abundance $\Omega_\chi h^2$, as soon as $m_\chi \gtrsim \Trh$, as is clear from Eq.~\eqref{Eq:omega3}.
Indeed, compensating the Boltzmann suppression with stronger couplings induces a rapid growth of $\sigmae$ for $m_\chi \gtrsim \Trh$. 
Since both the DM production rate and $\sigmae$ scale as $(\epsilon g'')^2$, increasing $\epsilon g''$ from $\sim 10^{-11}$ to $\sim 10^{-9}$ suffices to bring the predicted cross sections within current experimental sensitivity. 
As a consequence, present experimental limits already constrain masses $m_\chi \gtrsim 200~\mathrm{MeV}$ for $\Trh\gtrsim 1~$GeV. 
The projected reach of DAMIC-M and PandaX, shown as shaded regions in Fig.~\ref{fig:xs_lowTrh}, is observed to intersect a sizable portion of parameter space consistent with FISC production for $10~\mathrm{MeV} \lesssim \Trh \lesssim 10~\mathrm{GeV}$. 
As a first conclusion, we see that relaxing the assumption of high reheating temperature therefore substantially enlarges the phenomenologically testable parameter space.

\subsection{Constraint on $\Trh$}

Conversely, the explicit dependence of $\sigmae$ on the reheating temperature can be exploited to derive constraints on $\Trh$ as a function of the DM mass. 
For a fixed value of $m_\chi$ and the right amount of DM, experimental upper limits on $\sigmae$ translate, through Eq.~\eqref{Eq:omega3}, into bounds on the maximal enhancement of the couplings that can compensate for the Boltzmann suppression of DM production at low reheating temperatures. 
This, in turn, defines a critical reheating temperature below which the required couplings is excluded by direct-detection searches. Decreasing $\Trh$ further would require even larger couplings, eventually violating the fundamental assumption of freeze-in, namely that the dark sector never attains thermal equilibrium.

\begin{figure}[!t]
\centering
\includegraphics[width=0.5\textwidth]{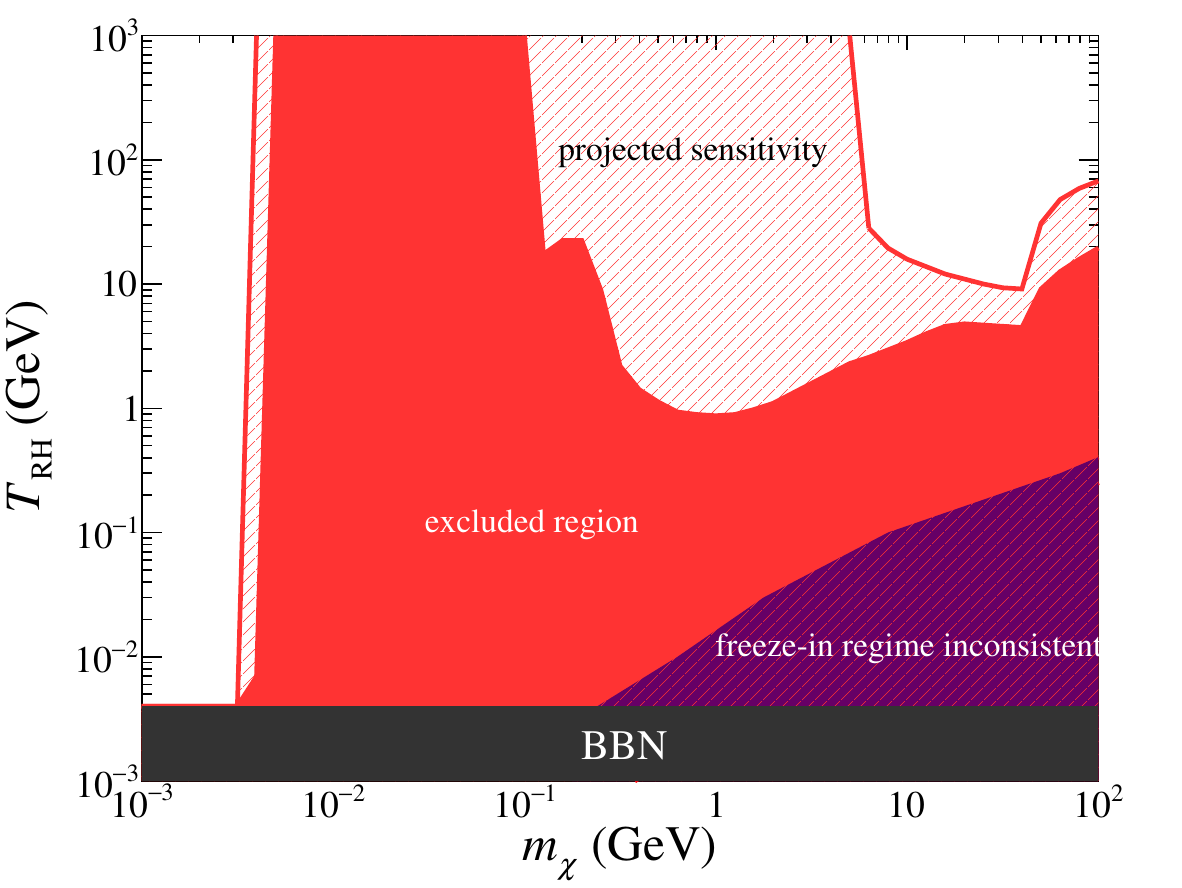} 
\caption{Exclusion zones of $\Trh$ as a function of $m_\chi$.}
\label{fig:Trh_vs_mX}
\end{figure}

Our results are displayed in Fig.~\ref{fig:Trh_vs_mX}, where we show the region of parameter space excluded by DAMIC-M in the $(m_\chi, T_{\rm RH})$ plane, under the requirement $\Omega_\chi h^2 = 0.12$. 
A notable feature is that DAMIC-M excludes a broad band in $T_{\rm RH}$ for $5~\mathrm{MeV} \lesssim m_\chi \lesssim 100~\mathrm{MeV}$. 
This behavior can be understood from the transition between production regimes. As $m_\chi \lesssim T_{\rm RH}$, the system departs from the FISC regime and enters the standard freeze-in (FIMP) regime, which is infrared dominated. 
In this case, DM production is controlled by temperatures $T \sim m_\chi$, and corresponds to retaining only the first term in Eq.~\eqref{Eq:omegatotal}. 
One then finds $\Omega_\chi h^2 \propto \sigmae$, implying $\sigmae \simeq 3\times 10^{-37}\,\mathrm{cm}^2$ to reproduce the observed relic abundance, see Eq.~\eqref{Eq:omega3}. 
This explains the convergence of iso-relic-density contours in Fig.~\ref{fig:Trh_vs_mX} for $m_\chi \lesssim \Trh$.

For larger masses, extending up to the electroweak scale, DAMIC-M constraints can in fact become more stringent than those derived from BBN. 
In particular, reheating temperatures as low as $T_{\rm RH} \sim 2~\mathrm{MeV}$ are incompatible with $m_\chi \gtrsim 5~\mathrm{MeV}$ within the FISC framework, unless one allows for values of $\sigmae$ that exceed current experimental bounds so as to compensate for the Boltzmann suppression of the production rate. 
In this sense, direct-detection experiments such as DAMIC-M provide a novel probe of the early Universe, effectively constraining the reheating temperature. 
As illustrated in Fig.~\ref{fig:Trh_vs_mX}, these bounds can reach $\Trh \gtrsim \mathcal{O}(10~\mathrm{GeV})$ for electroweak-scale DM.

It is also worth noting that the constraint on $\Trh$ becomes increasingly stringent with increasing $m_\chi$, approximately following $\Trh \propto m_\chi$. 
This scaling can be understood from the behavior of the interaction rate: for large $m_\chi$, one has $\sigmae \to \mathrm{const.}$, see Eq.~\eqref{Eq:sigma_e}, while $n_\chi \propto 1/m_\chi$, leading to an overall suppression of the production rate. 
From Eq.~(\ref{Eq:omega3}), this corresponds to the condition $\exp\left(2 m_\chi / T_{\rm RH}\right) \sim \mathrm{const.}$, and hence $T_{\rm RH} \propto m_\chi$, in agreement with the behavior observed in Fig.~\ref{fig:Trh_vs_mX}.

The projected sensitivity of the DAMIC-M detector is displayed as the shaded area. 
The accessible parameter space extends to reheating temperatures above $\sim 50~\mathrm{GeV}$ for electroweak-scale DM. 
This further highlights the potential of direct detection experiments as probes of reheating and early-Universe cosmology  within the FISC framework.

Finally, we also display in dark purple the parameters space where the freeze-in mechanism is inconsistent in the regime of stronger couplings due to the thermalization of the particles that turns out to occur in the dark sector. 
Determining this region requires identifying the coupling strengths for which the dark sector remains out of thermal equilibrium throughout its cosmological evolution. This condition can be expressed as
\begin{equation}
\Gamma(\Tds(T))<H(T),
\end{equation}
where $\Tds(T)$ denotes the temperature that the dark sector would have if it were in thermal equilibrium. 
To compute it, we introduce the dark-sector energy density, denoted as $\rhods$. 
At thermal equilibrium, and assuming first, to fix the ideas, that $\Tds\gg m_\chi$ (and $\Tds\gg m_{A'}$, which is satisfied given that we consider ultra-light $A'$ throughout this work), $\rhods$ would be related to the dark-sector temperature by
\begin{equation}
\label{Eq:rho_ds}
\rhods(\Tds)=\frac{g_{\star}^\ast\pi^2}{30}{\Tds}^4,
\end{equation}
with $g^\ast_{\star}$ denoting the effective number of relativistic degrees of freedom in the dark sector that amounts to $\frac{11}{2}$ for $m_{A'}=0$, or $\frac{13}{2}$ for $m_{A'}>0$. 
The rate of change of the would-be thermal equilibrium $\rhods$ is fueled by the rate of energy density $\mathcal{S}_{\rhods}(t)$ transferred from the SM sector to the dark one,
\begin{equation}
\label{Eq:boltzmann_rhods}
\frac{\d\rhods}{\d t}+4H\rhods=\mathcal{S}_{\rhods}(t).
\end{equation}
The back-reactions transferring energy from the dark sector to the SM one remain negligible as long as $\Tds$ does not approach $T$. 
Solving Eq.~\eqref{Eq:boltzmann_rhods} (see Appendix~\ref{app:numerical}) established a relationship between the temperature of the SM thermal bath and the dark sector one. 
The interaction rate that brings the dark sector to kinetic equilibrium can then be calculated as
\begin{equation}
\label{Eq:rate_DS}
\Gamma(\Tds)=n_\chi^\mathrm{eq}(\Tds) \langle\sigma v\rangle_{\chi},
\end{equation}
from which upper bounds on $\epsilon g''$ can be derived to ensure the self-consistency of the freeze-in regime. 
Scanning over $m_\chi$, we then obtain the exclusion region in the $(m_\chi,T_\mathrm{rh})$ plane shown in Fig.~\ref{fig:Trh_vs_mX}. 
The lower-right portion of the figure corresponds to values of $\epsilon g''$ large enough to thermalize the dark sector during reheating, thereby invalidating the freeze-in assumption.

\section{Inflaton-seeded hybrid freeze-in}
\label{sec:infdecay}

\subsection{Motivations}

So far, we have assumed that no initial production of DM occurred before the onset of the freeze-in process. 
This assumption is, however, highly questionable. 
Even though FISC-type models allow the weakness of the effective interaction rate to be attributed to Boltzmann suppression rather than to extremely feeble couplings, they do not resolve the question of preexisting DM production. 
Indeed, it has been shown in Refs.~\cite{Mambrini:2021zpp,Clery:2022wib,Clery:2021bwz,Henrich:2024rux,Barman:2024kyw,Barman:2022qgt,Lyth:1996yj,Chung:1998zb,Kuzmin:1999zk,Chung:2001cb,Kolb:2023ydq} that gravitational  production, both from the thermal bath and inflaton scattering through graviton exchange {\it after} inflation, can significantly contribute to the DM population,  independently of any subsequent production. 
This component should therefore, in principle, be included as an initial condition in Eq.~(\ref{Eq:Xlowrh}). 
In addition, it has recently been demonstrated that fluctuations of spectator (or inert) scalar fields generated during inflation can significantly populate the Universe once these modes re-enter the horizon~\cite{Choi:2024bdn,Garcia:2025rut}. 
More generally, unless one assumes a specific inflaton structure--for instance a ``dark-matter-phobic'' inflaton suppressing direct decays into dark matter--it is difficult to construct a fully consistent reheating scenario without some amount of early DM production, even with the introduction of reheaton fields. 
The only generic mechanism capable of erasing such initial conditions is the establishment of thermal equilibrium, allowing DM to redistribute thermally its degrees of freedom. 
This is the case for WIMPs in the non-relativistic regime, or for UFO scenarios \cite{Henrich:2025gsd,Henrich:2025sli,Henrich:2025pca} in the relativistic case. 

Although sufficiently low reheating temperatures--as required in FISC scenarios with sub-electroweak-scale DM--can strongly suppress gravitational production rates, which scale as $\propto T^8/\Mp^4$~\cite{Mambrini:2021zpp,Garny:2017kha}, it nevertheless remains difficult to justify a negligible preexisting DM abundance in the absence of thermalization. 
On the other hand, DM may also be produced directly or indirectly through inflaton decays~\cite{Kaneta:2019zgw,Garcia:2020eof,Bernal:2021qrl,Dudas:2017rpa,BhupalDev:2013oiy}.
In this section, we therefore compute the DM density produced {\it between} the end of inflation {\it and} the onset of radiation domination at $\Trh$ due to the decay of the inflaton $\phi$. 
Indeed, $\Trh$ corresponds to the temperature of the moment at which the inflaton has transferred its energy density to the thermal bath so that the Universe becomes radiation dominated. 
Consequently, inflaton decays no longer contribute directly to DM production  {\it after} reheating, but instead provide an initial condition for the subsequent freeze-in evolution beginning at $T_{\rm RH}$.
Throughout the remainder of this work, we consider that a quadratic inflaton potential dominates the end of the reheating process, 
$V(\phi)=\frac12 m_\phi^2 \phi^2$, 
assuming perturbative inflaton decays into either fermions or bosons~\cite{Garcia:2020eof,Garcia:2020wiy,Garcia:2021iag}. 
The inflaton decay width is then constant.
Additional details regarding the reheating dynamics are provided in Appendix~\ref{app:numerical}.

\subsection{The inflaton  decay}

Denoting the inflaton width as $\Gamma_\phi$, its mass as $m_\phi$ and its energy density as $\rho_\phi$, and assuming a branching ratio $B_{\phi\chi\chi}$ for a decay channel into a pair of $\chi$, the new source term of DM production in Eq.~\eqref{Eq:boltzmannt} reads as
\begin{equation}\label{Eq:S_phi}
    R_{\phi}(T)=\frac{2B_{\phi\chi\chi}\Gamma_\phi}{m_\phi}\rho_\phi(T)\,,
\end{equation}

\noi
where the factor of 2 comes from the fact 2 DM particles are produced per decay. For a quadratic potential of the inflaton, the energy density of the early Universe is dominated by $\rho_\phi(a)$, which behaves like dust (see Appendix~\ref{app:numerical}) for details),
\begin{equation}\label{Eq:rho_phi}
    \rho_\phi(a)=\rho_{\phi,\mathrm{RH}}\left(\frac{\arh}{a}\right)^3\,.
\end{equation}

\noi
Adopting the convention for the reheating scale factor to signal the equality between the radiation and inflaton energy densities, the constant $\rho_{\phi,\mathrm{RH}}$ reads as
\begin{equation}
    \rho_{\phi,\mathrm{RH}}=\frac{g_{\star \rho}(\Trh)\pi^2}{30}\Trh^4 =\rhorh \,,
\end{equation} 

\noi
where $g_{\star \rho}(T)$ stands for the effective degrees of freedom for energy density.
Using Eqs.~\eqref{Eq:T} and~\eqref{Eq:H} during reheating, the expression of $\rho_\phi$ as a function of the temperature is then
\begin{equation}\label{Eq:rho_phi_bis}
    \rho_\phi(T)=\frac{\grh\pi^2}{30}\left(\frac{g_{\star\rho}(T)}{\grh}\right)^2\frac{T^8}{\Trh^4}\,,
\end{equation}

\noi
with $\grh \equiv g_{\star \rho}(\Trh)$.

On the other hand, the width of the inflaton, which is related to the Hubble rate at reheating as $\Gamma_\phi\simeq \frac{3 \tilde c}{2}\Hrh$, with $ \tilde c \simeq1.2$ a constant obtained from a 
numerical integration~\cite{Ellis:2015jpg,Pradler:2006hh}, can be expressed as
\begin{equation}
    \label{Eq:Gamma_phi}
    \Gamma_\phi=\frac{\sqrt{3} \tilde c}{2}\left(\frac{\grh\pi^2}{30}\right)^{\frac{1}{2}}\frac{\Trh^2}{\Mp}\,.
\end{equation}

\noi 
Eq.~(\ref{Eq:boltzmanna}) 
{\it before} reheating then becomes\footnote{There also exists a contribution from the scattering of thermal bath particles during the reheating phase. However, since this contribution is of second order, we will not consider it further in this work.} 

\begin{equation}
    \frac{d X_\chi^\phi}{d a}=\frac{2 a^2}{H(a)} \Bphi \gamma_\phi
    \frac{\sqrt{\rhorh}}{M_P}\frac{\rho_\phi(a)}{m_\phi}\,,
    \label{Eq:xphi1}
\end{equation}

\noindent
with $\gamma_\phi=\frac{\sqrt{3}}{2} \tilde c\simeq 1.04$.
Using Eq.~\eqref{Eq:rho_phi}, Eq.~\eqref{Eq:xphi1} then becomes:

\begin{equation}
\frac{dX_\chi^\phi}{da} = 2 a^2 \Bphi \gamma_\phi \sqrt{3} \frac{\rhorh}{m_\phi} \left(\frac{\arh}{a}\right)^{3/2}.
\label{Eq:xphi2}
\end{equation}

\noi
It is then sufficient to integrate Eq.~\eqref{Eq:xphi2} from the end of inflation, at $a = \aend$, to reheating at $a = \arh$, assuming that any production {\it during} inflation has been sufficiently diluted ($X_\chi^\phi(\aend) = 0$). We then obtain:

\begin{equation}
X_\chi^\phi(a) = \frac{4 \Bphi \gamma_\phi}{\sqrt{3}} \frac{\rhorh}{m_\phi} \left( \arh^{3/2} a^{3/2} - \arh^{3/2} \aend^{3/2} \right)\,.
\end{equation}

\noi
Neglecting $\aend \ll \arh$, this yields to

\begin{equation}
n_\chi^\phi(a) \simeq \frac{4}{\sqrt{3}} \Bphi \gamma_\phi \frac{\rhorh}{m_\phi} \left( \frac{\arh}{a} \right)^{3/2}\,.
\end{equation}

\noi
Applying Eq.~\eqref{Eq:omegageneric}, we then deduce

\begin{multline}
\frac{\Omega_\chi^\phi h^2}{0.12}\simeq \left(\frac{\Bphi}{10^{-5}}\right)\left(\frac{\Trh}{100~\rm MeV}\right)
\left(\frac{m_\chi}{1~\rm GeV}\right)\\
\times \left(\frac{4.1 \times 10^3~\rm GeV}{m_\phi}\right).
\label{Eq:omegaphi}
\end{multline}

Note that we do not fix $m_\phi$ to a typical mass scale for Starobinsky-like models of inflation, where $m_\phi \simeq 2\times 10^{13}$ GeV. 
This approach is motivated by our aim to remain as generic as possible, assuming that the quadratic potential
$V(\phi) = \frac{1}{2} m_\phi^2 \phi^2$
dominates the end of the reheating process.
In particular, we do not specify the \textit{nature} of the field $\phi$ (which could be a reheaton), nor the full form of the potential.
Indeed, it was shown in Ref.~\cite{Clery:2024dlk} that even in models with an initially quartic potential, $V(\phi)\propto \lambda \phi^4$, a quadratic term, or a bare mass term, $\propto m_\phi^2\phi^2/2$, can always dominate the late stages of reheating due to the redshift evolution $\phi \propto a^{-1}$ at the end of inflation. 
Furthermore, mixed potentials containing both quadratic and quartic terms may provide additional phenomenological advantages, for instance in stabilizing the Higgs potential, as discussed in Ref.~\cite{Cado:2025orb}.
For these reasons, and to remain as model independent as possible, we treat $m_\phi$ as a free parameter for the remainder of this study.
It is also worth noting, from Eq.~\eqref{Eq:omegaphi}, that a small branching ratio $B_{\phi\chi\chi}$ can be compensated by a smaller inflaton mass $m_\phi$. 
This is because reducing $m_\phi$ increases the inflaton lifetime, thereby extending the duration of reheating, which defines the reheating temperature $\Trh$. 
Since a smaller reheating temperature leads to reduced dilution of the DM abundance, the observed relic density can then be recovered with a correspondingly smaller value of the branching ratio.

\subsection{Analysis}

We are now in a position to determine the DM relic density by combining Eqs.~\eqref{Eq:omegatotal} and~\eqref{Eq:omegaphi}, taking into account its production during reheating. This is given by:

\begin{multline}
\frac{\Omega_\chi}{0.12} \simeq  3.75\times 10^{22} |{\cal M}|^2 \left(1 + 2 \frac{m_\chi}{T_{\rm RH}} \left(\frac{g_\star}{g_{\rm RH}}\right)^{1/3}\right) \ee^{-\frac{2m_\chi}{\Trh}} \\ 
 + \left(\frac{\Bphi}{10^{-5}}\right) \left(\frac{\Trh}{100~\rm MeV}\right) \left(\frac{m_\chi}{1~\rm GeV}\right) \left(\frac{4.1 \times 10^3~\rm GeV}{m_\phi}\right).\nonumber
\end{multline}

which gives, using Eq.~\eqref{Eq:omega3},
\begin{multline}
\frac{\Omega_\chi}{0.12} \simeq  \left(1+2\left(\frac{100~\rm MeV}{\Trh} \right)\left(\frac{m_\chi}{1~\rm GeV}\right)\right)\\
\times  \left(\frac{\bar{\sigma}_e}{3 \times 10^{-37}~\rm cm^2}\right)
\ee^{-\frac{2m_\chi}{T_{\rm RH}}} \\
+ \left(\frac{\Bphi}{10^{-5}}\right) \left(\frac{\Trh}{100~\rm MeV}\right) \left(\frac{m_\chi}{1~\rm GeV}\right) \left(\frac{4.1 \times 10^3~\rm GeV}{m_\phi}\right)\,,
\end{multline}

\noi 
or, for $m_\chi \gg \Trh$,


\begin{multline}
\frac{\Omega_\chi}{0.12} \simeq  \left(\frac{100~\rm MeV}{\Trh} \right)\left(\frac{m_\chi}{1~\rm GeV}\right)\\
\times \left(\frac{\bar{\sigma}_e}{1.5 \times 10^{-37}~\rm cm^2}\right)
\ee^{-\frac{2m_\chi}{T_{\rm RH}}} \\
+ \left(\frac{\Bphi}{10^{-5}}\right) \left(\frac{\Trh}{100~\rm MeV}\right) \left(\frac{m_\chi}{1~\rm GeV}\right) \left(\frac{4.1 \times 10^3~\rm GeV}{m_\phi}\right).
\label{Eq:master}
\end{multline}

\noindent
Inverting the relation, we obtain 


\begin{multline}
\frac{\bar{\sigma}_e}{1.5 \times 10^{-37}~\rm cm^2}\simeq  \bigg[\left(\frac{\Omega_\chi}{0.12} \right)\left(\frac{\Trh}{100~\rm MeV} \right)\left(\frac{1~\rm GeV}{m_\chi}\right)\\
-\left(\frac{\Bphi}{10^{-5}}\right) \left(\frac{\Trh}{100~\rm MeV}\right)^2 \left(\frac{4.1 \times 10^3~\rm GeV}{m_\phi}\right)\bigg]\ee^{\frac{2 m_\chi}{\Trh}}\,.
\label{Eq:sigmatotal}
\end{multline}

\begin{figure}[!ht]
\centering
\includegraphics[width=0.49\textwidth]{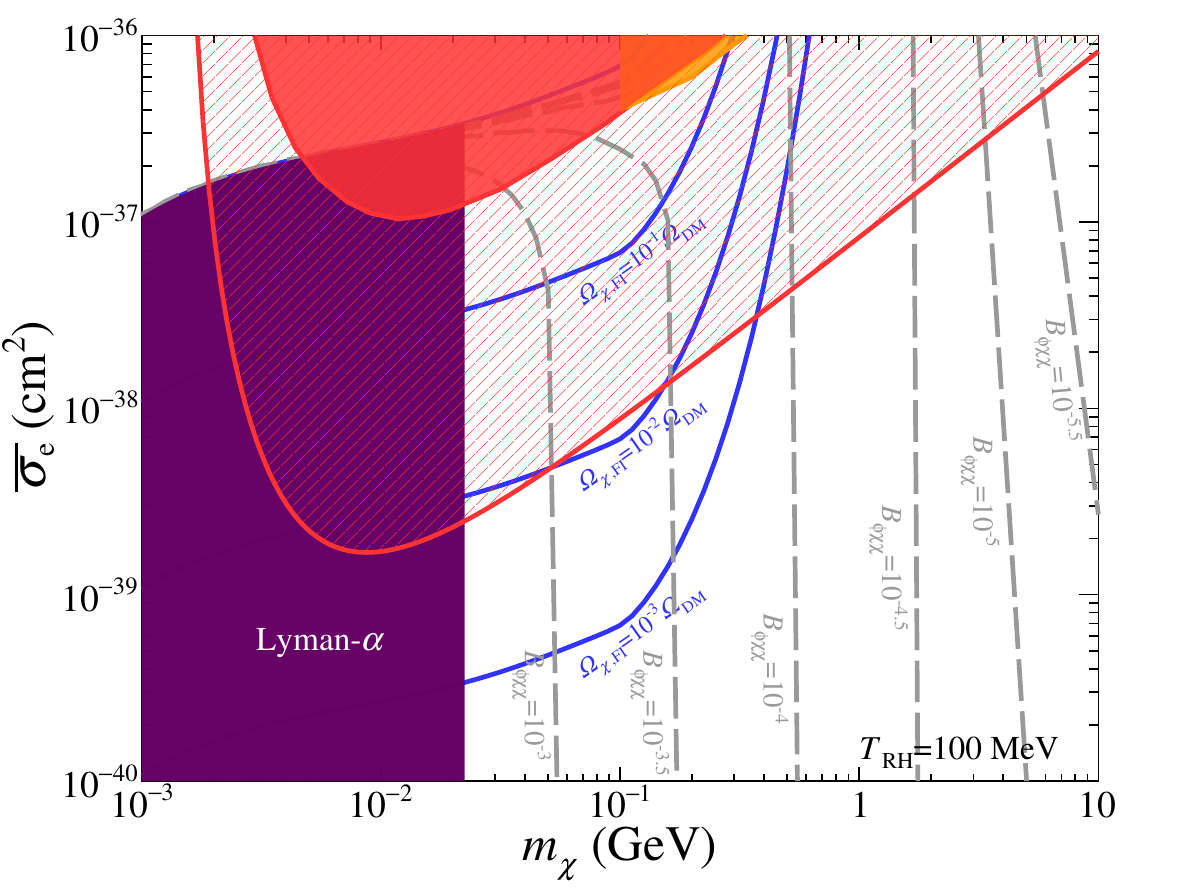}
\includegraphics[width=0.49\textwidth]{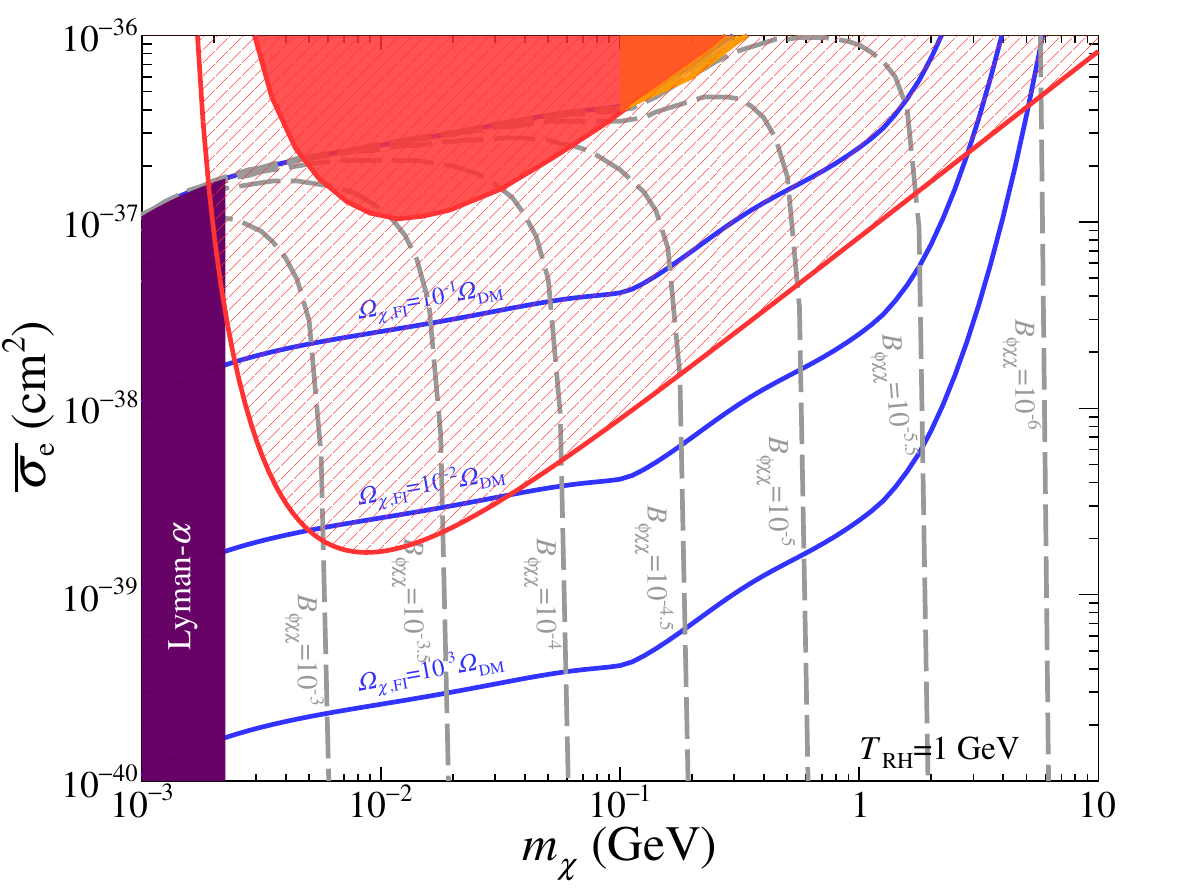}
\caption{\label{fig:sigma-B} Contours (in dashed gray) of the branching ratio of the inflaton decay into DM as a function of $\sigmae$ and $m_\chi$ to match a fraction $(1-f)$ of the relic density of DM observed today. The curves of $\sigmae$ set by freeze-in correspond to $f=(1,10^{-1},10^{-2},10^{-3})$ of the relic density. Top: $\Trh=100\,$MeV. Bottom: $\Trh=1\,$GeV.}
\end{figure}

In Fig.~\ref{fig:sigma-B}, we show as the gray dashed lines the iso-branching ratio curves in the $(m_\chi, \sigmae)$ plane. 
For each iso-curve, we clearly observe a sharp drop at a certain value of $m_\chi$. 
This corresponds to the threshold above which the production \textit{during} reheating becomes dominant. 
This behavior can be understood as follows. 
The production of DM from the decay of the inflaton does not depend on $m_\chi$, but only on its coupling to the inflaton parametrized here by the branching ratio $\Bphi$. 
In contrast, the FISC-type production from the thermal bath is exponentially suppressed according to $\ee^{-\frac{2m_\chi}{\Trh}}$, as shown in Eq.~\eqref{Eq:omega2}. 
Consequently, freeze-in production from the thermal bath is heavily suppressed above a sufficiently large DM mass. 
In this regime, the relic density is no longer controlled by the dark matter--visible sector coupling encoded in $\sigmae$, but instead entirely by the dark matter--inflaton coupling; this explains why the iso-$\Bphi$ contours become effectively independent of $\sigmae$ beyond the transition mass visible in Fig.~\ref{fig:sigma-B}.
From Eq.~\eqref{Eq:sigmatotal}, we see that this corresponds to

\beq
\frac{m_\chi}{1\,\rm GeV} \simeq \left(\frac{10^{-5}}{\Bphi}\right)\left(\frac{100~\rm MeV}{\Trh}\right)
\left(\frac{m_\phi}{4.1 \times 10^3~\rm GeV}\right)\,,
\eeq

\noi
which is effectively what we observe in Fig.~\ref{fig:sigma-B}.
In this case, the dependence of the relic abundance to the direct detection cross section $\sigmae$ obviously disappears.

An interesting alternative is to fix the fraction of DM $f$ produced by the decay of the inflaton relative to the total density. 
This is not equivalent to fixing the branching ratio, since the density from FISC strongly depends on $m_\chi$, unlike the density from inflaton decay. 
In other words, fixing $\Bphi$ as we did in Eq.~\eqref{Eq:sigmatotal} is {\it not} equivalent to fixing $f$, where $f$ is defined by

\begin{equation}
f = \frac{\Omega^{\rm FI}_\chi}{\Omega^{tot}_\chi}
\label{Eq:f}\,,
\end{equation}

\noi
where $\Omega^{tot}_\chi = \Omega^{\rm FI}_\chi + \Omega^\phi_\chi = 0.12$, 
\noi
By combining Eq.~\eqref{Eq:f} with Eq.~\eqref{Eq:omega3}, we obtain

\begin{equation}
\frac{\bar \sigma_e}{2.3 \times 10^{-37}\rm{cm^2}} \simeq  \, f \times \left(\frac{\Trh}{100~\rm MeV}\right) \left(\frac{1~\rm GeV}{m_\chi}\right) e^{\frac{2 m_\chi}{\Trh}} \, \,,
\end{equation}

\noi
We thus see that the dependence on the fraction of DM from freeze-in is simply a linear rescaling of $f$ from the result obtained in the previous section. 
This is also effectively what we observe as the blue lines in Fig.~\ref{fig:sigma-B}. 
This effect is a fairly logical consequence, since $f$ directly links the direct detection rate to the portion of the relic density produced solely by freeze-in.

Finally, we also display in dark purple exclusion regions from structure formation bounds, in particular those inferred from measurements of the Lyman-$\alpha$ forest~\cite{Bernal:2021qrl}. 
The key feature here is that DM particles produced directly from inflaton decay are highly relativistic at their creation, with a typical initial momentum $p_\mathrm{in}\simeq m_\phi/2$. 
Because these particles are highly energetic, they can free-stream over cosmological distances before becoming non-relativistic. 
If their free-streaming length is too large, the growth of density perturbations at small scales gets suppressed. 
Such a suppression would leave an observable imprint on the matter power spectrum, which is tightly constrained by Lyman-$\alpha$ forest data through the absorption features in the spectra of distant quasars. 
As most of the DM is produced when the scale factor is close to  $\arh$, the present-day momentum is obtained by redshifting the initial value,
\begin{equation}
    \label{Eq:p0}
    p_0 = \frac{a_{\rm rh}}{a_0}\frac{m_\phi}{2}, 
\end{equation}
which yields to
\begin{equation}
    \label{Eq:p0}
    \frac{p_0}{1\,\rm GeV}
    \simeq 4.5\times 10^{-10}\left(\frac{m_\phi}{10^3\,\rm GeV}\right)\left(\frac{100\,\rm MeV}{\Trh}\right).
\end{equation}
Lyman-$\alpha$ measurements are commonly interpreted as setting a lower bound on the mass of thermal warm dark matter, typically $m_\mathrm{WDM}\geq 3.5~$keV~\cite{Irsic:2017ixq}, which corresponds to an upper bound on the present-day velocity $v_\chi\leq 1.8\times 10^{-8}$~\cite{Masina:2020xhk}. Requiring that the velocity of the non-thermally produced $\chi$ particles today, $v_\chi\simeq p_0/m_\chi$, does not exceed this limit yields a lower bound on $m_\chi$:
\begin{equation}
    \label{Eq:mX_vs_Trh}
    \frac{m_\chi}{1\,\rm GeV}\gtrsim 2\times10^{-2} \left(\frac{m_\phi}{10^3\,\rm GeV}\right)\left(\frac{100\,\rm MeV}{\Trh}\right),
\end{equation}
which explains the dark purple areas in Fig.~\ref{fig:sigma-B}.

\section{Conclusion}
\label{sec:conclusion}

Motivated by recent exclusion limits obtained using data of the DAMIC-M and PandaX direct-detection experiments, which rule out the ``vanilla'' freeze-in production mechanism for DM masses in the range $3~\mathrm{MeV} \lesssim m_\chi \lesssim 1~\mathrm{GeV}$ under the assumptions of ultra-light mediator and high reheating temperatures, we have investigated theoretical uncertainties inherent to the freeze-in framework. 
In particular, we have identified model-dependent effects--namely, variations in the reheating temperature and contributions from additional production channels (e.g. inflaton or reheaton decay)--that can significantly alter the mapping between the DM relic abundance and the expected electron-scattering cross section, $\sigmae$. 
We have demonstrated that, within this broader parameter space, viable scenarios persist in which the relic abundance satisfied the \textit{Planck} constraints, while yielding $\sigmae$ values well below current experimental limits. 
These results underscore the importance of accounting for cosmological and model-building uncertainties when interpreting experimental constraints on freeze-in DM.

One of the uncertainty stems from the value of the reheating temperature. Allowing $\Trh$ to lie below the electroweak scale, down to the BBN limit, opens a Boltzmann-suppressed regime where viable relic abundance can still be achieved through enhanced couplings. 
For $m_\chi \gtrsim \Trh$, this regime is known as \textit{freeze in at stronger coupling}, or FISC. 
It can lead to larger $\sigmae$ values, improving prospects for direct detection in some mass range. 
Notably, for $\Trh \gtrsim 5~\mathrm{GeV}$, the projected sensitivity of the final DAMIC-M exposure intersects a substantial portion of the viable FISC parameter space that would be otherwise out of reach in the vanilla freeze-in framework, as one can see in Fig.~\ref{fig:xs_lowTrh}. 
These results highlight the critical impact of reheating temperature assumptions on freeze-in phenomenology and demonstrate that even secluded models may offer promising detection prospects in the near future.

Besides, we have considered the combined contributions of freeze-in and inflaton decay to the DM yield and mapped out the viable parameter space in $(\sigmae, m_\chi,\Trh, \Bphi)$ consistent with the relic abundance constraint.
For this scenario, we considered a generic inflationary potential with a bare mass term, i.e. $V(\phi) \sim \lambda \phi^4 + \frac{1}{2} m_\phi^2 \phi^2$. 
In this case, reheating occurs at a scale of the order of $m_\phi$, which we left as a free parameter. 
We noted that for $m_\phi$ of the order of $10^3$ GeV, a good relic density with $m_\chi = 1$ GeV can be obtained for a decay branching ratio $\Bphi \sim 10^{-5}$ and $\Trh \sim 100$ MeV, as we see in 
Eq.~\eqref{Eq:omegaphi}. 
Lower $m_\phi$ masses (or higher reheating temperatures) require smaller decay widths $\phi \rightarrow \chi \chi$ to avoid overproduction of DM during the reheating period. 
This scenario is well represented in Fig.~\ref{fig:sigma-B}.

For $\Trh$ of the order of a few hundred of MeV,  obtaining the right relic density for $m_\chi < 100~\mathrm{GeV}$ calls for a branching ratio of order of $B_{\phi \chi \chi} \in [10^{-6},10^{-3}]$. 
This analysis highlights the complementarity between the freeze-in and inflaton-decay contributions and emphasizes the possibility for direct-detection experiments to probe cosmological parameters in addition to the particle-physics ones pertaining to the gauge extensions of the SM.
It should also be noted that, while our study has been conducted within 
the framework of an extra $U_X(1)$ extension, only the calculation of the amplitude $|{\cal M}|^2$ is truly necessary for our analysis. 
Consequently, our framework can easily be applied to any other extension of the Standard Model.

In other words, the main objective of this work was to illustrate how the current absence of signal in direct-detection experiments such as DAMIC-M and Panda-X can provide additional motivation for investigating reheating dynamics and nonstandard DM production mechanisms.
In scenarios involving FISC-type scenarios, characterized by $m_\chi \gtrsim T_{\rm RH}$, as well as in scenarios where DM receives contributions from external sources such as inflaton decays, improved direct-detection limits translate directly into constraints on reheating-era parameters. 
Depending on the dominant production mechanism, these constraints may apply either to the reheating temperature itself, as shown in Fig.~\ref{fig:Trh_vs_mX}, or to the inflaton branching ratio into dark matter, $\phi \rightarrow \chi\chi$, as illustrated in Fig.~\ref{fig:sigma-B}.
Direct detection experiments thus become
cosmic explorers.

\begin{acknowledgments}

The authors would like to thank Stephen Heinrich and Keith Olive for
helpful discussions. 
This project has received funding from the European Union’s Horizon Europe research and innovation programme under the Marie Skłodowska-Curie Staff Exchange grant agreement No 101086085 – ASYMMETRY and the CNRS-IRP project UCMN.
We also acknowledge support by Institut Pascal at Université 
Paris-Saclay during the Paris-Saclay Astroparticle Symposium 2025.

\end{acknowledgments}

\appendix

\section{$U(1)$ extensions of electroweak interactions with kinetic mixing}
\label{app:models_withepsilon}

To make canonical the kinetic sector of the Lagrangian in the presence of a nonzero kinetic mixing, we have used Eq~\eqref{Eq:nonhat2hat} to transform the fields. Actually, the most general  transformation in terms of orthonormal fields $(\hat{W}_{3\mu},\hat{B}_{\mu},\hat{C}_{\mu})$ reads as
\begin{equation}
    \begin{pmatrix}
        W_{3\mu}\\
        B_\mu\\
        C_\mu
    \end{pmatrix}=\begin{pmatrix}
        1 & 0 & 0 \\
        0 & \gamma_\epsilon\cos\theta & -\gamma_\epsilon\sin\theta\\
        0 & \sin\theta-\epsilon\gamma_\epsilon\cos\theta & \cos\theta+\epsilon\gamma_\epsilon\sin\theta        
    \end{pmatrix} \begin{pmatrix}
        \hat{W}_{3\mu}\\
        \hat{B}_\mu\\
        \hat{C}_\mu
    \end{pmatrix},
\end{equation}
with $W_{3\mu}$ the neutral component of $SU(2)_\L$ generators,  $\gamma_\epsilon=(1-\epsilon^2)^{-1/2}$, and $\theta$ an angle that remains free as long as the extra-$U(1)_\X$ symmetry is unbroken. 

The setup considered throughout the paper corresponds to $\sin\theta=\epsilon$. In this case, the covariant derivative takes, to first order in $\epsilon$, the form
\begin{equation}\label{Eq:Dmu-g''eps}
    D_\mu=\partial_\mu-\mathrm{i}gT_3\hat{W}_{3\mu}-\frac{\mathrm{i}}{2}g'Y\hat{B}_\mu
    -\frac{\mathrm{i}}{2}\left(g''Q_X-\epsilon g'Y\right)\hat{C}_\mu,
\end{equation}
so that $\hat{C}_\mu$ gets coupled to SM particles through their hypercharge, with an effective couplings $\epsilon g'$. Introducing one single SM Higgs doublet with charges $T_3=-1/2$ and $Y=1$, the gauge-boson mass sector of the Lagrangian, after electroweak symmetry breaking, reads as
\begin{multline}\label{Eq:Lm}
    \mathcal{L}_\mathrm{m}=-\frac{1}{2}\frac{v^2}{4}\bigg(g^2\hat{W}_{3\mu}\hat{W}_{3}^{\mu}+g'^2\hat{B}_{\mu}\hat{B}^{\mu}+\epsilon^2g'^2\hat{C}_{\mu}\hat{C}^{\mu}\\
    -2gg'\hat{W}_{3\mu}\hat{B}^{\mu}+2\epsilon gg' \hat{W}_{3\mu}\hat{C}^{\mu}-2\epsilon g'^2\hat{B}_\mu \hat{C}^\mu\bigg),
\end{multline}
with $v$ the vacuum expectation value of the Higgs. The three physical gauge bosons are the eigenstates of the corresponding mass matrix; they are obtained from the hatted fields by $3\times 3$ rotations around $\hat{B}_\mu$ and $\hat{C}_\mu$:
\begin{equation}\label{Eq:AZA'}
    \begin{pmatrix}
        \tilde{Z}_\mu\\A_\mu\\A'_\mu
    \end{pmatrix}=\mathcal{R}_{\hat{B}_\mu}(+\xi)\cdot \mathcal{R}_{\hat{C}_\mu}(+\thetaW)\begin{pmatrix}
        \hat{W}_{3\mu}\\\hat{B}_\mu\\\hat{C}_\mu
    \end{pmatrix},
\end{equation}
with $\thetaW$ the electroweak mixing angle and $\xi\simeq\epsilon\sin\thetaW$. In this particular electroweak extension, the $A_\mu$ boson is the standard photon and the $A_\mu'$ boson is the ``dark photon''. The third physical gauge bosons is a modified version of the SM $Z_\mu$ boson, with a mass receiving corrections to order $\epsilon^2$. Expressing the covariant derivative in terms of the physical gauge bosons, it is straightforward to infer the coupling constants between each gauge boson and any particle with quantum numbers $(T_3,Y,X)$. To first order in $\epsilon$, they read as
\begin{eqnarray}\label{Eq:g_setup1}
    g_{\tilde{Z}}(T_3,Y,Q_X)&=& \frac{e}{\sW\cW}(T_3-\s2W Q)+ \epsilon g''\sW \frac{Q_X}{2},  \\
    g_{A}(T_3,Y)&=& g\sW T_3+g'\cW \frac{Y}{2} (\equiv eQ), \\
    g_{A'}(T_3,Y,Q_X)&=&-\epsilon e\cW Q+g'' \frac{Q_X}{2}.
\end{eqnarray}
Here, $\cW$ ($\sW$) [$\tW$] stands for $\cos\thetaW$ ($\sin\thetaW$) [$\tan\thetaW$]. With this particular choice of $\theta$ angle, the photon remains blind to the dark-sector particles while the dark photon couples to both the dark sector with couplings $g''$ and to SM particles with electroweak couplings reduced by $\epsilon$. In addition, the $\tilde{Z}$ boson inherits from a couplings $\epsilon g''$ to DM compared to its SM version.  The cross sections of interest can then be calculated by plugging these coupling constants in the expressions given in Appendix~\ref{app:xs}. They are proportional to $(\epsilon g'')^2$.

Another possible choice is to consider $\theta=0$ (referred to as ``Holdom phase''). This choice is complementary to $\sin{\theta}=\epsilon$ in the sense that it reverses the roles of the photon and the dark photon. The covariant derivative takes now the form
\begin{equation}\label{Eq:Dmu-eps2}
    D_\mu=\partial_\mu-\mathrm{i}gT_3\hat{W}_{3\mu}-\frac{\mathrm{i}}{2}(g' Y-g''\epsilon  Q_X)\hat{B}_\mu
    -\frac{\mathrm{i}}{2} g''Q_X\hat{C}_\mu
\end{equation}
to first order in $\epsilon$. The physical gauge bosons are obtained from the hatted fields by a single $3\times 3$ rotation around $\hat{C}_\mu$:
\begin{equation}\label{Eq:AZA'}
    \begin{pmatrix}
        Z_\mu\\A_\mu\\A'_\mu
    \end{pmatrix}=\mathcal{R}_{\hat{C}_\mu}(+\thetaW)\begin{pmatrix}
        \hat{W}_{3\mu}\\\hat{B}_\mu\\\hat{C}_\mu
    \end{pmatrix}.
\end{equation}
In other words, the three bosons are the photon $A_\mu$, the SM $Z_\mu$ and the dark photon $A'_\mu$ identified with the $C_\mu$. The coupling constants read as
\begin{eqnarray}\label{Eq:g_setup2}
    g_{Z}(T_3,Y,Q_X)&=&\frac{e}{\sW\cW}(T_3-\s2W Q)+\epsilon g'' \sW \frac{Q_X}{2},  \\
    g_{A}(T_3,Y,Q_X)&=& eQ - \epsilon g'' \cW \frac{Q_X}{2}, \\
    g_{A'}(Q_X)&=&g'' \frac{Q_X}{2}.
\end{eqnarray}
As anticipated, the situation is  reversed compared to the previous setup: the dark photon is blind to SM particles while both the photon and the $Z_\mu$ couple to the dark-sector particles with a ``milli-charge'' $\epsilon g''$.  

It is understood that any other choice of angle $\theta$ would ``interpolate'' between the two extreme choices made here: both the photon and the dark photon would be coupled to both the dark and SM sectors.

\section{Squared matrix elements}
\label{app:xs}

For completeness, we provide in this appendix the full expressions of $\int\d\Omega\overline{\left|\mathcal{M}\right|^2}$ for DM production from thermal-bath particles in the limit of massless dark photon:

\begin{widetext}
    \begin{itemize}
        \item Symmetric EW phase, $T>T_\mathrm{EW}$:
        \begin{equation}
        \int\d\Omega\overline{\left|\mathcal{M}\right|^2}= \frac{\pi}{6} \epsilon^2 g'^2g''^2(Y_{f\mathrm{L}}^2+Y_{f\mathrm{R}}^2)n_\mathrm{c}\left(1+\frac{2m_\chi^2}{s}\right),
        \end{equation}
        with $Y_f$ the chiral hypercharge of SM fermions. 
        \item Broken EW phase, $T<T_\mathrm{EW}$:
        \begin{multline}
        \int\d\Omega\overline{\left|\mathcal{M}\right|^2}= \frac{4\pi}{3}(\epsilon eg'')^2n_\mathrm{c}\Bigg[Q_f^2\cW^2\left(1+\frac{2m_f^2}{s}\right)\left(1+\frac{2m_\chi^2}{s}\right)-Q_fv_f\frac{s(s-m_Z^2)}{(s-m_Z^2)^2+(\Gamma_Zm_Z)^2}\\
        +\left(\frac{v_f^2}{4\cW^2}\left(1+\frac{2m_f^2}{s}\right)\left(1+\frac{2m_\chi^2}{s}\right)+\frac{a_f^2}{4\cW^2}\left(1-\frac{4m_f^2}{s}\right)\left(1+\frac{2m_\chi^2}{s}\right)\right)\frac{s^2}{(s-m_Z^2)^2+(\Gamma_Zm_Z)^2}\Bigg]
        \end{multline}
        with $m_f$ the mass of SM fermions, $v_f=T_3-2Q\sW^2$ the vector couplings of the $\tilde{Z}$ and $a_f=T_3$ the axial one, and $n_\mathrm{c}$ the number of colors of SM fermions. 

        
    \end{itemize}
\end{widetext}

\section{Numerical integration of Boltzmann equation}
\label{app:numerical}

In this appendix, we provide the details to solve numerically Eqs.~\eqref{Eq:boltzmannt} and.~\eqref{Eq:boltzmann_rhods}. To do so, we consider separately the two distinct eras of reheating and radiation. 

The radiation era is standard in cosmology, with $T\propto 1/a$. On the other hand, the reheating era takes place right after the period of exponential expansion in a matter-dominated background of classic harmonic oscillations for the inflaton about a minimum of potential. We consider here a quadratic potential about its minimum. As the inflaton decays, its by-products thermalize and the temperature of the primordial plasma rises quickly to a maximum, $\Tmax$.  Subsequently, the temperature falls as $T\propto a^{\sfrac{-3}{8}}$, with $a$ is the cosmological scale factor, until the Universe enters into a radiation-dominated era at $\Trh$~\cite{Giudice:2000ex}. Solving the reheating dynamics by accounting for the temperature dependence of the thermalized energy density, the temperature can be parameterized as~\cite{Garcia:2020wiy,Garcia:2020eof}
\begin{equation}\label{Eq:T}
    \frac{T(a)}{\Trh}=\left\{
    \begin{array}{ll}
        \left(\frac{\arh}{a}\right)^{\sfrac{3}{8}}\left(\frac{g_{\star \rho}(\Trh)}{g_{\star \rho}(T)}\right)^{\sfrac{1}{4}}  \mbox{if } \amax\leq a < \arh, \\
         \left(\frac{\arh}{a}\right)\left(\frac{g_{\star s}(\Trh)}{g_{\star s}(T)}\right)^{\sfrac{1}{3}}  \mbox{if } a\geq\arh,
    \end{array}
\right.
\end{equation}
with $\amax=(8/3)^{2/5}\aend$ the scale factor at the maximum temperature $\Tmax$ and $\aend$ that at the end of inflation. Hereafter, $g_\star(T)$ ($g_{\star \rho}(T)$) [$g_{\star s}(T)$] stand for the effective degrees of freedom for number (energy) [entropy] density. The reheating temperature, $\Trh$, signals the beginning of the radiation era.  Under these conditions, the Hubble rate evolves as
\begin{equation}
    \label{Eq:H}
    \frac{H(a)}{\Hrh}= \left\{
    \begin{array}{ll}
        \left(\frac{\arh}{a}\right)^{\sfrac{3}{2}}  \mbox{ if } \amax\leq a < \arh, \\
         \left(\frac{g_{\star \rho}(T)}{g_{\star \rho}(\Trh)}\right)^{\sfrac{1}{2}}\left(\frac{g_{\ast s}(\Trh)}{g_{\ast s}(T)}\right)^{\sfrac{2}{3}}\left(\frac{\arh}{a}\right)^{2}  \mbox{ if } a\geq\arh,
    \end{array}
\right.
\end{equation} 
where $g_\star(T)$ stands for the relativistic number of degrees of freedom at temperature $T$ and $\Hrh$ for the Hubble scale at the end of the reheating era,
\begin{equation}\label{Eq:Hrh}
    \Hrh\equiv H(\arh)=\left(\frac{g_{\star \rho}(\Trh)\pi^2}{90}\right)^{\sfrac{1}{2}}\frac{\Trh^2}{\Mp}.
\end{equation}

During the reheating era, Eq.~\eqref{Eq:boltzmannt} can be readily re-expressed in terms of the auxiliary function $\mathcal{Y}_\chi$ defined as $\mathcal{Y}_\chi=n_\chi/\tilde{T}^8$, with the auxiliary temperature $\tilde{T}=(g_{\star \rho}(T))^{\sfrac{1}{4}}T$. In this way, the term stemming from the expansion of the universe gets absorbed and the rate of change of $\mathcal{Y}_\chi$ reads as
\begin{equation}
    \label{Eq:Ychi-rh}
    \frac{\d \mathcal{Y}_\chi}{\d t}=\frac{R(t)}{\tilde{T}^8},
\end{equation}
which, using $\d t=\sfrac{-8\d\tilde{T}}{3H\tilde{T}}$ and the relation between $T$ and $\tilde{T}$, leads to
\begin{equation}
    \label{Eq:Ychi-rh-bis}
    \mathcal{Y}_\chi(\Trh)=\int_{\Trh}^{\Tmax}\d T\frac{8\tilde{g}_\rho(T)R(T)}{3g_{\star\rho}^2(T)HT^9},
\end{equation}
with $\tilde{g}_\rho(T)=1+(\sfrac{T}{4})(\sfrac{\d\log{g_{\star \rho}(T)}}{\d T})$ and where we used $\mathcal{Y}_\chi(\Tmax)=0$.

During the radiation era, on the other hand, we consider this time the auxiliary function $\mathcal{X}_\chi$ defined as $\mathcal{X}_\chi=n_\chi/\hat{T}^3$, with $\hat{T}=(g_{\star s}(T))^{\sfrac{1}{3}}T$. In this way, the rate of change of $\mathcal{X}_\chi$ reads as
\begin{equation}
    \label{Eq:Ychi-rad}
    \frac{\d \mathcal{X}_\chi}{\d t}=\frac{R(t)}{\hat{T}^3},
\end{equation}
which, using $\d t=\sfrac{-\d\hat{T}}{H\hat{T}}$, leads to
\begin{equation}
    \label{Eq:Ychi-rad-bis}
    \frac{\d \mathcal{X}_\chi}{\d T}=-\frac{\tilde{g}_s(T)R(T)}{g_{\star s}(T)HT^4},
\end{equation}
with $\tilde{g}_s(T)=1+(\sfrac{T}{3})(\sfrac{\d\log{g_{\star s}(T)}}{\d T})$. The DM energy density observed today is then
\begin{equation}
    \label{Eq:nrh}
    \rho_{\chi,0}=m_\chi g_{\star s}(T_0)T_0^3\left(\int_{T_0}^{\Trh}\d T\frac{\tilde{g}_s(T)R(T)}{g_{\star s}(T)HT^4}+\mathcal{X}_{\chi,\mathrm{rh}}\right),
\end{equation}
with $\mathcal{X}_{\chi,\mathrm{rh}}=\mathcal{Y}_{\chi}(\Trh)\Trh^5(\sfrac{g_{\star \rho}^2(\Trh)}{g_{\star s}(\Trh)})$.

The energy density $\rho_{\chi,0}$ is customarily expressed relative to the critical energy $\rho_\mathrm{c}=3.61\times 10^{-47}~h^2~\mathrm{GeV}^{4}$ to yield to the dimensionless relic abundance
\begin{equation}\label{Eq:OmegaX}
    \Omega_\chi h^2=\frac{\rho_{\chi,0}}{\rho_\mathrm{c}/h^2},
\end{equation}
with $h=H_0/(100~\mathrm{km\,s^{-1}\,Mpc^{-1}})$. To match the relic abundance of DM $\Omega_\mathrm{DM} h^2$ as measured with \textit{Planck}, we impose then $\Omega_\chi h^2\simeq0.12$. When integrating Eqs.~\eqref{Eq:Ychi-rh-bis} and \eqref{Eq:nrh}, the source term must be treated differently depending on whether the temperature lies above or below the electroweak transition temperature, $T_\mathrm{EW}\simeq 149.5$ GeV. For $T>T_\mathrm{EW}$, the source term is fed exclusively by annihilations of thermal-bath particles $i$~\cite{Gondolo:1990dk},
\begin{equation}
    R_{T>T_\mathrm{EW}}(T)=
    \sum_f~(n_f^\mathrm{eq})^2\langle\sigma v\rangle_{ff\rightarrow\chi\chi},
\end{equation}
with $n_f^\mathrm{eq}(T)$ the number density of SM fermions $f$ at temperature $T$ and $\langle\sigma v\rangle_{ff\rightarrow\chi\chi}$ the thermal-averaged cross sections of interest. Overall, and using the shorthand notation $\beta_i(x)=\sqrt{1-\sfrac{4m_i^2}{x}}$, the term reads as
\begin{multline}
    \label{Eq:S_aboveTEW}        
    R_{T>T_\mathrm{EW}}(T)=\frac{T}{4(2\pi)^6}\int\d s \sqrt{s}\beta_\chi(s) \mathcal{K}_1\left(\sfrac{\sqrt{s}}{T}\right) \\
    \times \sum_f\int\d \Omega\overline{\left|\mathcal{M}\right|^2}_{ff\rightarrow\chi\chi},
\end{multline}
where the statistical kernel $\mathcal{K}_1(x)$ that accounts for Fermi-Dirac statistics for relativistic fermions is customarily approximated as $\sfrac{3}{4}\times K_1(x)$~\cite{Dolgov:1992wf}, with $K_1(x)$ the modified Bessel function of second kind, and where we accounted for a factor $2$ for the creation of 2 DM particles in each annihilation process. The squared-matrix element is detailed in Appendix \ref{app:xs}. On the other hand, when the temperature drops below $T_\mathrm{EW}$, the source term is modified to account for the mass of the DM particles and receives an additional contribution from the $\tilde{Z}_\mu$ decay in models where it couples to DM,
\begin{multline}
    \label{Eq:S_belowTEW}        
    R_{T<T_\mathrm{EW}}(T)=\frac{T}{4(2\pi)^6}\int\d s \sqrt{s}\beta_\chi(s)K_1\left(\sfrac{\sqrt{s}}{T}\right)
    \\ 
    \times 
    \sum_f\beta_f(s)\int\d \Omega\overline{\left|\mathcal{M}\right|^2}_{ff\rightarrow\chi\chi}\\
    +\frac{(g''\epsilon \sW)^2}{24\pi} n^\mathrm{eq}_{\tilde{Z}}(T)m_Z  \beta_\chi(m_Z^2)\left(1+\frac{2m_\chi^2}{m_Z^2}\right),
\end{multline}
with $n^\mathrm{eq}_{\tilde{Z}}(T)$ the equilibrium function describing the density distribution of $\tilde{Z}$ particles at temperature $T$.

Eq.~\eqref{Eq:boltzmann_rhods} can be solved using the same strategy, making the distinction between the reheating and radiation eras. During the reheating era, considering the auxiliary quantity $\mathcal{Y}^\dagger_\rho=\rhods/\tilde{T}^n$, the Boltzmann equation reads as
\begin{equation}
    -\frac{3}{8}H\tilde{T}^{n+1}\frac{\d \mathcal{Y}^\dagger_\rho}{\d \tilde{T}}-\frac{3n}{8}H\tilde{T}^n\mathcal{Y}^\dagger_\rho+4H\tilde{T}^n\mathcal{Y}^\dagger_\rho=\mathcal{S}_{\rhods}(\tilde{T}).
\end{equation}
It is thus apparent that for $n=\sfrac{32}{3}$, the terms relative to the expansion of the Universe cancel. It is then straightforward to integrate the equation for the auxiliary variable and then to find
\begin{equation}
    \label{Eq:rho'-rh}
    \rhods(T)=g_{\star \rho}^{\sfrac{32}{12}}(T)T^{\sfrac{32}{3}}\int_T^{\Tmax}\d T'\frac{8\tilde{g}_\rho(T')\mathcal{S}_{\rhods}(T')}{3g_{\star \rho}^{\sfrac{32}{12}}(T')HT'^{\sfrac{35}{3}}}.
\end{equation}
On the other hand, during the radiation era, considering the auxiliary quantity $\mathcal{X}^\dagger_\rho=\rhods/\hat{T}^4$, $\rhods$ reads as
\begin{equation}
    \label{Eq:rho'-rad}
    \rhods(T)=g_{\star s}^{\sfrac{4}{3}}(T)T^4\left(\int_T^{\Trh}\d T'\frac{\tilde{g}_s(T')\mathcal{S}_{\rhods}(T')}{g_{\star s}^{\sfrac{4}{3}}(T')HT'^5}+\mathcal{X}_{\rho,\mathrm{rh}}\right).
\end{equation}
Requiring Eq.~\eqref{Eq:rho_ds} to be satisfied yields the dark-sector temperature as a function of $T$. The source term is dominantly fueled by processes leading to the production of DM particles with respect to that of dark photons, although the scaling as $(\epsilon g'')^2$ of the former reactions is \textit{a priori} less favorable than that of annihilation or coalescence, semi-Compton. and Bremsstrahlung processes ($\propto \epsilon^2$). The reason is the behavior of a kinetically mixed dark photon at finite temperature. In a thermal plasma, the ordinary photon acquires an effective in-medium mass set by the plasma frequency $\omega_\mathrm{p}(T)$, while the dark photon does not. After diagonalizing the kinetic and mass terms in this environment, the effective mixing angle between the propagating photon and dark-photon modes becomes temperature-dependent and scales parametrically as $\epsilon \left(\sfrac{m_{A'}}{\omega_\mathrm{p}}\right)^2$ for $m_{A'}\ll \omega_\mathrm{p}$. Consequently, in the nearly massless limit, the in-medium mixing angle vanishes, even if the vacuum kinetic mixing parameter is nonzero. Since on-shell dark-photon production proceeds through this effective mixing, the corresponding production rate is suppressed by a factor $m_{A'}^4$. On the other hand, off-shell exchanges do not require the use of in-medium eigenstates so that the source term is~\cite{Chu:2011be}
\begin{multline}
    \label{Eq:sigmavE}
    \mathcal{S}_{\rhods}(T)=\frac{T}{4(2\pi)^6}\int\d s~ \sqrt{s}\beta_\chi(s) K_2\left(\sfrac{\sqrt{s}}{T}\right)\\ 
    \times \sum_f\int\d \Omega\overline{\left|\mathcal{M}\right|^2}_ {ff\rightarrow\chi\chi}.
\end{multline}
during the reheating era. Modifications alongside those of Eq.~\eqref{Eq:S_belowTEW} hold in the same way here during the radiation era.

\nocite{*}

\bibliography{apssamp}

\end{document}